\documentclass[
    aps,                
    pra,                
    reprint,            
    superscriptaddress, 
    amsmath,            
    amssymb,            
    nofootinbib,        
    longbibliography
]{revtex4-2}
\usepackage[french,english]{babel}
\usepackage{graphicx}  
\usepackage{dcolumn}   
\usepackage{bm}        
\usepackage[colorlinks=true,allcolors=blue]{hyperref}  
\usepackage{upgreek}   

\usepackage[version=4]{mhchem}
\usepackage{amsmath}
\usepackage{amssymb}
\usepackage{tabularx}
\usepackage{dsfont}
\usepackage{tensor}
\usepackage{physics}
\usepackage{slashed}
\usepackage{simpler-wick}

\def\R{\mathbb{R}}

\def\C{\mathbb{C}}
\DeclareMathOperator{\sgn}{sgn}

\def\beq{\begin{equation}}
\def\eeq{\end{equation}}
\def\be*{\begin{equation*}}
\def\ee*{\end{equation*}}

\begin{document}

\title{Wichmann--Kroll vacuum polarization density in a finite Gaussian basis set}

\author{Ryan Benazzouk}
\email{ryan.benazzouk@irsamc.ups-tlse.fr}
\affiliation{Laboratoire de Chimie et Physique Quantique, Université de Toulouse, Toulouse, France}

\author{Maen Salman}
\email{maen.salman@lkb.upmc.fr}
\affiliation{Laboratoire Kastler Brossel, Sorbonne Université, CNRS,
ENS-Université PSL, Collège de France, 4 place Jussieu, F-75005 Paris, France}

\author{Trond Saue}
\email{trond.saue@irsamc.ups-tlse.fr}
\affiliation{Laboratoire de Chimie et Physique Quantique, Université de Toulouse, Toulouse, France}

\date{\today}

\begin{abstract}
This work further develops the calculation of QED effects in a finite Gaussian basis. We focus on the non-linear $\alpha(Z\alpha)^{n\ge 3}$ contribution to the vacuum polarization density, computing the energy shift of $1s_{1/2}$ states of hydrogen-like ions. Our goal is to improve the numerical computations to achieve a precision comparable to that of Green's function methods reported in the literature. To do so, an analytic expression for the linear contribution to the vacuum polarization density is derived using Riesz projectors. Alternative formulations of the vacuum polarization density and their relation is discussed. The convergence
of the finite Gaussian basis scheme is investigated, and the numerical difficulties that arise are characterized. In particular, an error analysis is performed to assess the method's robustness to numerical noise. We then report a strategy for computing the energy shift with sufficient precision to enable a sensible extrapolation of the partial-wave expansion. A key feature of the procedure is the use of even-tempered basis sets, allowing for an extrapolation towards the complete basis set limit.
\end{abstract}


\maketitle

\section{Introduction}

Vacuum polarization calculations have been the subject of theoretical developments for many decades. Recently, a new spark of interest arose with the search for physics beyond the Standard Model\cite{Safronova_RevModPhys.90.025008}. Very accurate tests of the Standard Model are often done in strong field experiments, either with few-electron high-Z ions or muonic atoms. In high-Z few-electron atoms, vacuum-polarization effects become comparable to the finite-nuclear-size correction \cite[Fig.2]{Johnson1985}, and will eventually overtake the electron self-energy, albeit at un realistically high nuclear charges\cite{Soff_PhysRevLett.48.1465,Thierfelder_PRA2010}. Moreover, in muonic atoms, vacuum polarization becomes the dominant QED contribution \cite{Borie_RevModPhys.54.67,Eides2007}, and its magnitude approaches, and can exceed, the finite-nuclear-size term as the nuclear charge increases \cite{CavusogluSikora2023}. For even heavier negatively charged bound particles, such as antiprotons, vacuum polarization remains a significant electromagnetic contribution at small radii \cite{Borie1983PhysRevA.28.555, PatkosPachucki2025, Sommerfeldt2025}. These observations underline the fundamental importance of vacuum polarization as a key probe of quantum electrodynamics in the strongest electromagnetic fields accessible in atomic systems.

Vacuum polarization was first introduced by Dirac \cite{Dirac:1934:TDP, Dirac:1934:DID} and Peierls \cite{Peierls1934}. Heisenberg \cite{Heisenberg_ZfP1934} identified the divergence to be removed in the first order of perturbation theory in the strength of the external potential, namely $\alpha (Z\alpha)$. This was then computed by Uehling \cite{Uehling}, followed by Serber \cite{Serber1935} as well as Pauli and Rose \cite{PauliRose}. Later Schwinger provided a different derivation of the vacuum polarization potential\cite{Schwinger1949II}. The first order of perturbation theory was found to be Uehling's result, and the logarithmic divergence of this contribution was more precisely handled by charge renormalization. Later, Karplus and Neuman showed how a logarithmic divergence present also in the $\alpha(Z\alpha)^3$ diagrams could be eliminated either by enforcing gauge-invariance in the calculations or by using Pauli--Villars regularization \cite{KarplusNeuman}.

Then came the seminal paper of Wichmann and Kroll \cite{WichmannKroll}, where they formulated the density of the polarized vacuum for an atom with a pointlike nucleus as a contour integral of the resolvent of the Dirac operator. They proved the density to be analytic, and used the Laplace transform of the radial Dirac Green's function to give an expression for the entire non-linear contribution, yielding the first non-perturbative result on vacuum polarization. Later research works largely adopted their formalism, as it proved to be the most suited for numerical approaches. When computing the Green's function numerically one can use its representation in terms of regular and irregular solutions \cite[Eq.~17]{WichmannKroll} for a direct evaluation of the integrals, or its spectral representation, which works best in finite basis methods \cite{Yerokhin2020}.

Blomqvist derived expressions for the contributions to the vacuum polarization potential, also for a pointlike nucleus, of orders $\alpha (Z\alpha)$, $\alpha^2 (Z\alpha)$ and $\alpha (Z\alpha)^3$ as well as approximations for $\alpha (Z\alpha)^5$ and $\alpha (Z\alpha)^7$, and provided a first evaluation of the energy shift of transition energies in munonic Pb \cite{Blomqvist1972}. Rinker and Wilets \cite{RinkerWilets1973, RinkerWilets1975}, followed by Gyulassy \cite{Gyulassy1974Muonic, Gyulassy1974Collisions, Gyulassy1975}, reported energy shift calculations in high-$Z$ ions, including supercritical atoms \cite{Gyulassy1974Collisions}, with more realistic nuclear models. Gyulassy notably demonstrated that the partial-wave expansion $\rho^{\text{WK}}=\sum_{\abs{\kappa}}\rho^{\text{WK}}_{\abs{\kappa}}$ removes the spurious logarithmic divergence in $\alpha(Z\alpha)^3$ term, yielding the finite physical contribution\cite[sec.2.2]{Gyulassy1975}. Similar calculations were performed by Neghabian in a momentum space approach \cite{Neghabian1983}. As noted by Neghabian, these calculations crucially show that even in the supercritical regime $Z\alpha > 1$, the Wichmann--Kroll contribution remains small and negative. Johnson and Soff provided a thorough comparison of contributions to the Lamb shift of hydrogen-like ions, including the $\alpha(Z\alpha)^3$ contribution and the higher orders $\alpha (Z\alpha)^5$ and $\alpha (Z\alpha)^7$ \cite{Johnson1985}. Non-perturbative calculations were undertaken by Soff and Mohr using the partial-wave method of Gyulassy, where they reported energy shift values for many high-$Z$ with finite nuclear models \cite{Soff_PRA1988}. They were later confirmed and expanded upon by results from Persson \textit{et al.}, who computed the vacuum-polarization potential directly using a partial-wave expansion, with the required set of bound and continuum wavefunctions obtained from a finite-difference solution of the Dirac equation in a spherical box, which enforces a discretization of the positive- and negative-energy continua \cite{Persson1993, Salomonson_PhysRevA.40.5548}. These results constitute the reference data for the present work. For the point-nucleus case non-perturbative calculations have been reported by Manakov and co-workers\cite{Manakov_JETP1989}, who in later works provided accurate numerical approximations to the non-linear contribution of both the potential and the density \cite{Fainshtein_JPhysB1991,Manakov_Vestnik2012,Manakov_Vestnik2013}.

Recently, attempts have been made at computing QED effects in finite Gaussian basis sets \cite{Salman_PRA2023, Salman_sym2020, Ferenc2025, Ivanov2024}. As noted by Yerokhin and Maiorova\cite{Yerokhin2020}, the finite basis approach benefits from both its simplicity of implementation and from the regularity features of the approximate Green's function. However, parameter dependencies such as the basis size or the number of partial waves included can limit the precision of the calculations, and so does the large cancellations that occur in the computations. We should add that linear dependencies in Gaussian basis sets represent an even greater obstacle for precision. Still, Salman and Saue showed that finite Gaussian basis sets can reproduce the non-linear contribution to the vacuum-polarization density with high accuracy \cite{Salman_PRA2023}. Ivanov \textit{et al.} subsequently evaluated the corresponding energy shifts and assessed their convergence by comparison with both B-spline calculations and reference results obtained from Green's-function methods \cite{Ivanov2024}. While the B-spline representation of the vacuum-polarization density exhibits strong oscillations, leading to sizeable uncertainties in the extracted energy shifts, the Gaussian basis provides numerically stable results for significantly smaller basis sizes. Further comparison with the high-precision results of Persson \textit{et al.} \cite{Persson1993} confirmed the convergence of the Gaussian-basis approach, albeit at the cost of linear-dependency issues that required expensive arbitrary-precision arithmetic. 

In the present work we intend to tackle all of these issues. Error bounds caused by linear dependencies are estimated, and show that quadruple precision is sufficient for any practical calculation of the energy shift in a hydrogen-like ions. In addition, the finite-basis limitations are dealt with using an extrapolation procedure. We report calculations  of the energy shift of ground states of a selection of high-$Z$ ions, comparing with Refs.\cite{Soff_PRA1988, Persson1993}.

Lastly, it is worth mentioning that this problem has also attracted increasing attention in the mathematics community. As such, an important source of inspiration has been the no-photon QED mean-field  model formulated by Chaix and Iracane \cite{Chaix1989I, Chaix1989II}. In this work the authors reformulate the problem of vacuum polarization in a variational setting. This was the motivation for a mathematically rigorous treatment of these calculations, including the renormalization procedure, in a series of articles initiated by Hainzl and Siedentop \cite{HainzlSiedentop, Barbaroux2005, Hainzl2005Existence, Hainzl2005Self, Hainzl2007, Gravejat2009, Gravejat2011, Lewin2010}. These results are of particular interest for us, as they formulate a many-potential expansion of the vacuum polarization density matrix. 

The outline of our paper is as follows: In Sec.~\ref{sec:theory} we provide the theory underlying our computational work. In Sec.~\ref{sec:defs} we discuss various formulations of vacuum polarization density and their relation. In Sec.~\ref{sec:riesz} we introduce Riesz projectors as a starting point for the definition of the Wichmann--Kroll energy shift in finite basis, discussed in Sec.~\ref{sec:basis}. In Sec.~\ref{sec:numerical_noise} we analyze numerical noise in our finite-basis calculation, principally arising from linear dependence, which is studied in detail in Sec.~\ref{sec:lindep}. Computational details are given in Sec.~\ref{sec:comp}. In Sec.~\ref{sec:results} we first present out procedure for the selection of exponents of our Gaussian basis functions, before we present the results of our calculations on hydrogen-like ions. We conclude and provide perspectives in Sec.~{sec:conc}.

All expressions used and developed in this work are written in SI units in order to facilitate their conversion to the favorite choice of units adopted by the reader.

\section{Theory}\label{sec:theory}

\subsection{Vacuum polarization for hydrogen-like ions in a finite basis}

The variational method \cite{Lewin2010} suggests a formulation of QED corrections that is akin to the mathematical apparatus of quantum chemistry, that is to compute any quantity in the basis of the eigenstates of the Dirac Hamiltonian. 
In what follows, we describe how the QED effects stemming from vacuum polarization can be formulated in this fashion, and how one can then compute them using a finite Gaussian basis approximation.

In this work we are only interested in time-independent radial potentials $\mathcal{A}_\mu(x)=\left(\frac{1}{c}\phi(r),\vec{0}\right)$. We work in the Furry picture, where the electronic field operator\cite{Mohr1998, Thaller}
\beq
	\hat{\psi}(x) = \sum_{\epsilon_n>-mc^2} \psi_n(x)\hat{a}_n + \sum_{\epsilon_m\leq-mc^2}\psi_m(x)\hat{b}_m^\dagger,
\eeq
is expressed in the generalized basis of solutions to the Dirac equation in the external potential
	\begin{eqnarray}\label{eq:Dirac_eq}
	\left[\beta mc^2-i\hbar c\vec{\alpha}\cdot\vec{\nabla}+V(r)\right]\psi_n(\vec{x}) &=& \epsilon_n\psi_n(\vec{x}) \\
	\psi_n(x) &=& \psi_n(\vec{x})e^{-\frac{i}{\hbar}\epsilon_n t}.
	\end{eqnarray}
We consider here one-electron ions, so the scalar potential energy can be expressed as
\beq
	V(r)= -\frac{Ze^2}{4\pi\epsilon_0}\int\frac{\nu(r^\prime)}{|\vec{r}-\vec{r}^{\prime}|}d^3r^\prime,
\eeq
where the normalized nuclear radial density $\nu$ is convoluted with the inverse distance to form a central potential and $e$ is the fundamental charge.

\begin{figure}[!h]
\centering
\includegraphics[width=\linewidth]{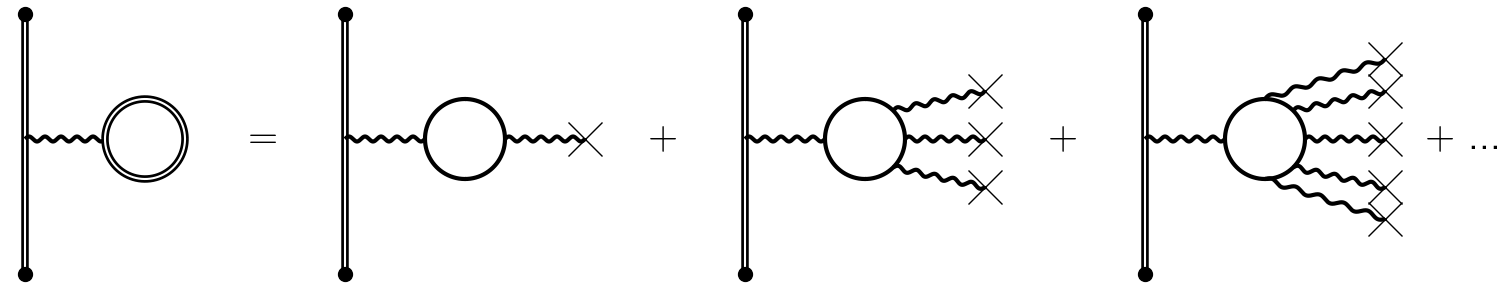}
\caption{Perturbative expansion in $Z\alpha$ of the vacuum polarization correction in the Furry picture.}
\label{fig:VP_diagrams}
\end{figure}

In the Furry picture electrons propagating through the external potential are represented diagrammatically by double lines, while simple lines represent the free electron propagators. We denote those propagators $S_\mathcal{A}$ and $S$, respectively. Here, we are interested in the vacuum polarization (VP) class of radiative loop correction diagrams, see Fig.\ref{fig:VP_diagrams}, corresponding to an instantaneous Coulomb interaction of the bound electron with a VP four-current
\beq \label{eq:VP_four_current}
	J_\mu^{\text{VP}}(x) = i\hbar ec\Tr\left[\gamma_\mu S_\mathcal{A}(x,x)\right].
\eeq
        
In the case of time-independent external potentials, and more specifically for a time-reversal symmetric Dirac Hamiltonian (vanishing three-vector potential), only the time-like component of the four-current contributes \cite{Salman_PRA2023}. This contribution corresponds to the vacuum polarization charge density \cite{Mohr1998}, which follows from Eq.~\eqref{eq:VP_four_current} and the spectral decomposition of $S_\mathcal{A}$,
\beq \label{eq:SchwingerVPdensity}
	\rho^{\text{VP}}(\vec{x}) = -\frac{e}{2}\left[\sum_{\epsilon_n\leq-mc^2}\abs{\psi_n(\vec{x})}^2 - \sum_{\epsilon_n>-mc^2}\abs{\psi_n(\vec{x})}^2\right].
\eeq
This expression of vacuum polarization is particularly interesting for our case, as it only involves densities of the Dirac Hamiltonian eigenstates. This can be straightforwardly implemented in a finite basis.

The diagrams of Fig.\ref{fig:VP_diagrams} translates into a series of corrections to the Coulomb potential energy \cite{Indelicato2014, Sommerfeldt2025}
\beq
	V^{\text{VP}}(r) = V^{\text{VP},(1)}(r)+V^{\text{VP},(3+)}(r),
\eeq
with $V^{\text{VP},(1)}$ being the Uehling potential \cite{Uehling}, and $V^{\text{VP},(3+)}$ the Wichmann--Kroll (WK) potential \cite{WichmannKroll}. The diagrammatic expansion is an expansion in the number of external potential vertices, that is an expansion in the external field strength $Z\alpha$. It is well-known that the linear term of this expansion, the Uehling potential $V^{\text{VP},(1)}$, is at first divergent and needs to be separately regularized. This poses a significant challenge for a finite basis approach. The Wichmann--Kroll potential however is finite, and can be very-well handled in a finite basis \cite{Salman_PRA2023, Ivanov2024}.

\subsection{Equivalent definitions of vacuum polarization} \label{sec:defs}

\begin{figure}[!h]
\centering
\includegraphics[width=\linewidth]{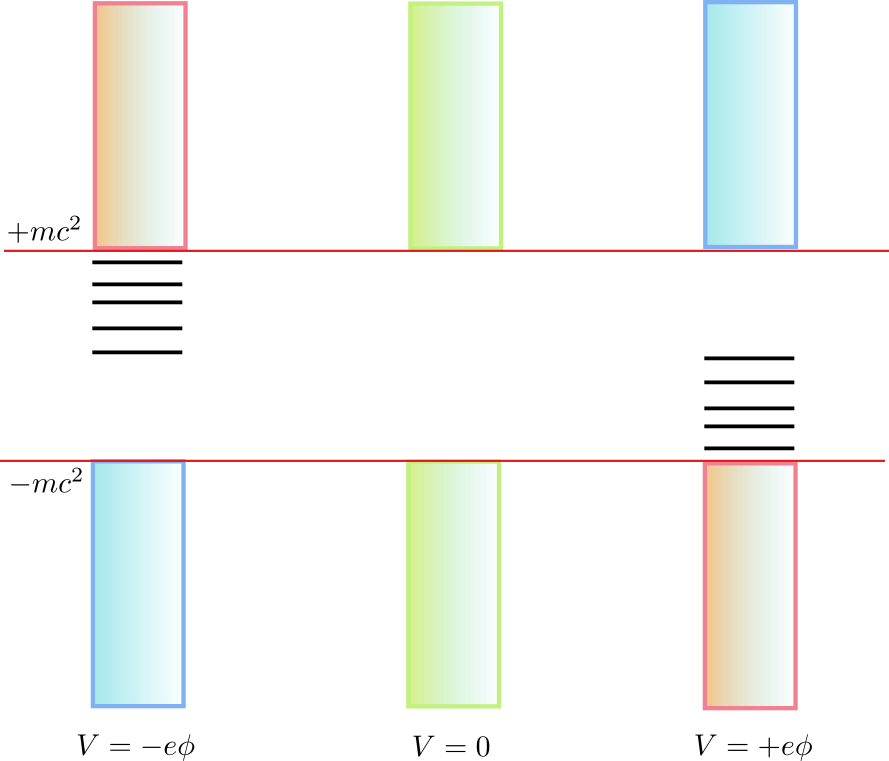}
\caption{Spectrum of Dirac equation with different choices of the potential energy term $V$.}
\label{fig:Dirac_spectrum}
\end{figure}

There is an alternative definition of vacuum polarization that stems directly from the Dirac sea picture of the QED vacuum,
\beq \label{eq:DiracVPdensity}
	\tilde{\rho}^{\text{VP}}(\vec{x}) = -e\left[\sum_{\epsilon_n\leq-mc^2}\abs{\psi_n(\vec{x})}^2 - \sum_{\epsilon_n^0\leq-mc^2}\abs{\psi_n^0(\vec{x})}^2\right].
\eeq
Here, $\{\psi_n^0\}$ are solutions to the free-particle Dirac equation, that is, setting  $V=0$ in Eq.~\eqref{eq:Dirac_eq}. This is in fact the first and quite
intuitive definition, given by Dirac in his report to the 7th Solvay Congress in Brussels in October 1933\cite{Dirac:1934:TDP}. The basic form of the now
conventional definition, Eq.~\eqref{eq:SchwingerVPdensity}, was also given by Dirac,
\cite{Dirac:1934:DID}, with a factor $1/2$ added by Heisenberg.\cite{Heisenberg_ZfP1934,Heisenberg_ZfP1934err} About this second definition, Dirac wrote: ``This has the advantage that it makes a closer symmetry between the electrons and the positrons and leads to neater mathematical expressions''.\cite{Dirac:1934:DID} It is hard to believe that Dirac did not see the
equivalence of these two definitions, but the earliest demonstration in print that we are aware of is that of Hamm and Sch{\"u}tte \cite{Hamm_JPhysA1990,Hamm:ddiplomarbeit} (see also Ref.~\cite{saue_2006talk}), using charge conjugation symmetry\cite{Kramers_1937} and completeness. We shall repeat the demonstration here, using a notation that allows us to explore further connections:

We rewrite Eq.~\eqref{eq:DiracVPdensity} as
\begin{equation} \label{DiracVP2}
\tilde{\rho}^{\text{VP}}=\rho_e^{(-)}-\rho_0^{(-)}=-e\left[n_e^{(-)} - n_0^{(-)}\right].
\end{equation}
We use the notation $n^{(\pm)}_{e/p}$ to refer to the \emph{number} density of solutions of positive (superscript $+$) or negative (superscript $-$) energy of the electronic (subscript $e$) or positronic (subscript $p$) problem, defined by the potential energy term $V=q\phi$ in the Dirac equation, Eq.~\eqref{eq:Dirac_eq}, setting charge $q=-e$ and $q=+e$ for electrons and positrons, respectively. The corresponding spectra are sketched in Fig.~\ref{fig:Dirac_spectrum}. For charge densities we use the notation $\rho^{\pm}_{e/p}$. No specification of charge is required in the free-particle problem, and we therefore use the subscript $0$. With this notation the conventional definition, Eq.~\eqref{eq:SchwingerVPdensity}, is expressed as
\begin{equation}
  \label{SchwingerVP2}
 \rho^{\text{VP}}(x)=\frac{1}{2}\left[\rho^{(-)}_e-\rho^{(+)}_e\right]=-\frac{e}{2}\left[n^{(-)}_e-n^{(+)}_e\right].
\end{equation}
We shall use the completeness relation
\begin{equation} \label{eq:completeness}
  n_{e}^{(+)}+n_{e}^{(-)}=n_{0}^{(+)}+n_{0}^{(-)}.
\end{equation}  
From charge conjugation, as inferred from Figure \ref{fig:Dirac_spectrum}, it follows that
\begin{equation}\label{chargeconj}
  n_{p}^{(\mp)}=n_{e}^{(\pm)}.
\end{equation}
In the free-particle case we have
\begin{equation}
  n_{0}^{(-)}=n_{0}^{(+)}\quad\Rightarrow\quad n_{e}^{(+)}+n_{e}^{(-)}=2n_{0}^{(-)},
\end{equation}
which immediately shows the connection between Eqs.~\eqref{DiracVP2} and ~\eqref{SchwingerVP2}. Furthermore, using Eq.~\eqref{chargeconj}, we can
rewrite Eq.~\eqref{SchwingerVP2} as 
\begin{equation}
  \label{SchwingerVP3}
 \rho^{\text{VP}}(x)=-\frac{e}{2}\left[n^{(-)}_e-n^{(-)}_p\right]=\frac{1}{2}\left[\rho^{(-)}_e+\rho^{(-)}_p\right],
\end{equation}
showing that it corresponds to taking the average of the electronic and positronic vacuum.

In the present work we focus on the vacuum polarization $\rho^{\text{VP}}(\vec{x},Z)$ of one-electron atoms of nuclear charge $Z$ calculated in finite basis sets.
An obvious extension of Eq.~\eqref{chargeconj} is
\begin{equation}
  \label{charge-conj2}
  n_p^{(\pm)}(\vec{x},+Z) = n_e^{(\pm)}(\vec{x},-Z).
\end{equation}
This leads to yet a reformulation of Eq.~\eqref{SchwingerVP2}
\begin{equation}
  \rho^{\text{VP}}(\vec{x},Z)=-\frac{e}{2}\left[n^{(-)}_e(\vec{x},Z)-n^{(-)}_e(\vec{x},-Z)\right].
\end{equation}
Upon a Taylor expansion in $Z$ one finds that even terms vanish, providing a simple demonstration of Furry's theorem\cite{Furry_PhysRev.51.125}.

For later use we note an alternative expression, using positive-energy solutions only, which is obtained starting from Eq.~\eqref{SchwingerVP2} and using Eq.~\eqref{chargeconj}
\begin{equation}\label{eq:VP-RKB}
  \rho^{\text{VP}}(\vec{x},Z)=-\frac{e}{2}\left[n^{(+)}_p(\vec{x},Z)-n^{(+)}_e(\vec{x},Z)\right].
\end{equation}

\subsection{Riesz projectors and the many-potential expansion}\label{sec:riesz}

All of the developments performed above hold perfectly well formally. However the charge density of the Dirac sea is not a well-defined quantity. For a more rigorous description of vacuum polarization, the quantity to be considered is the density matrix of the Dirac sea, as recognized by Dirac himself \cite{Dirac:1934:TDP, Dirac:1934:DID}. This picture makes easier the consideration of properties like the completeness relation seen in Eq.~\eqref{eq:completeness}, or the many-potential expansion (see below). To compute vacuum polarization in a finite basis, we have seen that special care must be taken when dealing with the linear term of the $Z\alpha$ expansion. We will now show that the consideration of density matrices makes possible the extraction of the linear term by an explicit computation of the many-potential expansion. \\
To do so, it is useful to follow the method of Ref.~\cite{HainzlSiedentop}. We should first recognize any density matrix as a spectral projector of the Dirac Hamiltonian. Let
\beq
	\hat{h}_0 = -i\hbar c\vec{\alpha}\cdot\vec{\nabla} + \beta mc^2
\eeq
be the free Dirac Hamiltonian defined on $L^2(\R^3,\C^4)$ with the domain $H^1(\R^3,\C^4)$ where it is self-adjoint \cite{Thaller}. In the presence of an external Coulomb potential energy $V$, the Coulomb-Dirac Hamiltonian $\hat{h} = \hat{h}_0 + V$ remains self-adjoint if $Z\alpha < \sqrt{3}/2$ \cite{Barbaroux2005}.
Following Dirac \cite{Dirac:1934:DID}, we can rewrite the sums in Eq.~\eqref{eq:DiracVPdensity} as the traces of the density matrices where all of the negative energy solutions are occupied, namely the Dirac sea. This corresponds to the spectral projector onto the negative energy eigenstates of the Dirac Hamiltonian \cite{Lewin2010}
\beq
	P_- = \chi_{(-\infty,-mc^2]}(\hat{h})\qcomma P_-(\vec{x},\vec{y}) = \sum_{E_n\le -mc^2}\phi_n(\vec{x})\phi_n^\dagger(\vec{y}),
\eeq
and the same goes for $P_-^0=\chi_{(-\infty,-mc^2]}(\hat{h}_0)$. These density matrices are associated with the formal densities $n_e^{(-)}$ and $n_0^{(-)}$, which are divergent. Nonetheless Eq.~\eqref{eq:DiracVPdensity} provides the form of a vacuum polarization density matrix $Q$ as
\beq \label{eq:dirac_vp_density_matrix}
	Q = P_- -P_-^0
\eeq
and the vacuum polarization density is given by
\beq \label{eq:vp_trace}
	\rho^{\text{VP}}(\vec{x}) = -e\Tr_{\C^4}Q(\vec{x},\vec{x}).
\eeq
This expression is still divergent, as it contains the $\alpha(Z\alpha)$ singularity. However, according to Lewin, it becomes well defined with the application of a momentum-space cutoff \cite{Lewin2010}. In this case, the vacuum polarization density matrix $Q$ is Hilbert-Schmidt and therefore has a kernel representation $Q(\vec{x},\vec{y})$ which is $L^2$ \cite[Th.VI.23]{ReedSimon}. In addition the operators $P^0_\pm Q P^0_\pm$ are trace-class, which makes the kinetic energy of the polarized vacuum $\Tr(|\hat{h}^0|Q)$ also finite \cite{Lewin2010}.

The spectral projectors can then be related to the resolvent of the Dirac Hamiltonian, in the most general case by Stone's formula \cite[Ch.6 Lemma 5.6 and Problem 5.7.]{Kato} (see also \cite[Th.VII.13]{ReedSimon} or \cite[Th. 4.19]{Lewin2022} for the case of a bounded self-adjoint operator, the proof of the latter generalizes straightforwardly to the unbounded case), or, in the case of a finite-dimensional basis set formulation, where the spectrum is necessarily discrete, using the Riesz projector \cite[Sec.3.6.5 Eq.~6.19]{Kato} \cite[Th. 4.18]{Lewin2022} \cite{Kato1949}. \\
In particular, we see that the density matrix $Q$ can be computed with a contour integral along the Feynman contour $\mathcal{C}_F$ as :
\beq \label{eq:vp_resolvent_integral}
	Q = \frac{1}{2i\pi}\int_{\mathcal{C}_F}\left(\frac{1}{\zeta - \hat{h}}-\frac{1}{\zeta-\hat{h}_0}\right)\mathrm{d}\zeta.
\eeq
Now if the nuclear density $\nu$ is not too strong \cite[Th. 3]{Hainzl2005Existence}, it can be considered a small perturbation to the free Dirac Hamiltonian $\hat{h} = \hat{h}_0 + V$, and we can use the many-potential expansion of the resolvent
\beq
	\frac{1}{\zeta-\hat{h}} = \sum_{N=0}^{+\infty}\frac{1}{\zeta-\hat{h}_0}\left(V\frac{1}{\zeta-\hat{h}_0}\right)^N.
\eeq
This provides a series expansion for $Q$ in powers of the external potential energy, or, in our case, in $Z\alpha$,
\beq \label{eq:vp_density_mat_expansion}
	Q = \sum_{N=1}^{+\infty}Q^{(N)},
\eeq
where $Q^{(1)}$ in particular has the form
\beq
	Q^{(1)} = \frac{1}{2i\pi}\int_{\mathcal{C}_F}\frac{1}{\zeta-\hat{h}_0}V\frac{1}{\zeta-\hat{h}_0}\mathrm{d}\zeta.
\eeq
Then, the linear contribution to the vacuum polarization density follows formally from the spectral representation of the resolvent $(\zeta-\hat{h}_0)^{-1}$ as :
\beq \label{eq:linvp_density_3d}
	\rho^{\text{VP},(1)}(\vec{x}) = -\frac{e}{2i\pi}\sum_{n,\ell}\int_{\mathcal{C}_F}\frac{\mel{\psi_n^0}{V}{\psi_\ell^0}}{(\zeta-\epsilon_n^0)(\zeta-\epsilon_\ell^0)}\psi_n^{0\dagger}(\vec{x})\psi_\ell^0(\vec{x})\mathrm{d}\zeta.
\eeq

Another characterization is worth mentioning at this point. In the case of an atomic potential, the expansion in Eq.~\eqref{eq:vp_density_mat_expansion} provides for the density a series expansion in powers of $Z\alpha$. For such a power series, the coefficients of the expansion follow as
\beq \label{eq:vp_finite_diff}
	\rho^{\text{VP},(N)}(\vec{x}) = \frac{Z^N}{N!} \left.\dv[N]{Z}\rho^{\text{VP}}(\vec{x})\right|_{Z=0}, 
\eeq
which was first noticed by Rinker and Wilets \cite{RinkerWilets1975}, and then used by Salman and Saue for their finite basis calculation of the Wichmann--Kroll density \cite{Salman_PRA2023}. \\

In a finite basis, we know that the spectrum of $\hat{h}$ and $\hat{h}_0$ will consist of a finite number of isolated eigenvalues. Then we can always find a closed contour $\mathcal{C}_-$ in the complex plane that circles all of the negative eigenvalues of both $\hat{h}$ and $\hat{h}^0$, so that Eq.~\eqref{eq:vp_resolvent_integral} becomes
\beq
	Q = \frac{1}{2i\pi}\oint_{\mathcal{C}_-}\left(\frac{1}{\zeta-\hat{h}}-\frac{1}{\zeta-\hat{h}_0}\right)\mathrm{d}\zeta.
\eeq
The linear contribution to the vacuum polarization density matrix follows likewise from Eq.~\eqref{eq:linvp_density_3d} and can be evaluated with the residue theorem as 
\beq
	Q^{(1)} = \sum_{\substack{\epsilon_n^0 > -mc^2 \\ \epsilon_\ell^0 \le -mc^2}}\left(\frac{\mel{\psi_n^0}{V}{\psi_\ell^0}}{\epsilon_n^0 - \epsilon_\ell^0}\dyad{\psi_n^0}{\psi_\ell^0}+\mbox{h.a.}\right),
\eeq
where h.a. denotes the Hermitian adjoint, and the linear contribution to the density follows as
\beq \label{eq:linvp_density_finite_basis}
	\rho^{\text{VP},(1)}(\vec{x}) = -2e\sum_{\substack{\epsilon_n^0 > -mc^2 \\ \epsilon_\ell^0 \le -mc^2}}\text{Re}\left(\frac{\mel{\psi_n^0}{V}{\psi_\ell^0}}{\epsilon_n^0 - \epsilon_\ell^0}\psi_n^{0\dagger}(\vec{x})\psi_\ell^0(\vec{x})\right).
\eeq

The density in Eq.~\eqref{eq:linvp_density_3d} holds only formally as this expression is divergent. Since Uehling \cite{Uehling}, it is known that it contains a finite physical contribution, and a non-physical quantity that must be removed by means of regularization and renormalization. In a finite basis there is no divergence. But Eq.~\eqref{eq:linvp_density_finite_basis} nonetheless contains a non-physical contribution that will diverge in the complete basis set limit. What can be straightforwardly studied in the finite basis set approximation is the Wichmann--Kroll contribution to the density
\beq
	\rho^{\text{WK}}(\vec{x}) = \rho^{\text{VP}}(\vec{x}) - \rho^{\text{VP},(1)}(\vec{x}).
\eeq
This can be evaluated along with its energy shift of any orbital $\psi^{\text{ref}}$ as
\beq \label{eq:energy_shift_def}
	\Delta E^{\text{WK}} = \frac{1}{4\pi\epsilon_0}\int\mathrm{d}^3x_1\int\mathrm{d}^3x_2\;\rho^{\text{ref}}(\vec{x}_1)\frac {1}{\abs{\vec{x}_1-\vec{x}_2}}\rho^{\text{WK}}(\vec{x}_2),
\eeq
with $\rho^{\text{ref}}(\vec{x})=-e\abs{\psi^{\text{ref}}(\vec{x})}^2$. In the atomic case, considered in this work the expression simplifies to
\beq\label{eq:atomic_VPshift}
\Delta E^{\text{WK}} = \frac{1}{4\pi\epsilon_0}(4\pi)^2\int_0^\infty dr_1r_1^2\int_0^\infty dr_2r_2^2\;\overline{\rho}_{\Omega}^{\text{ref}}(r_1)\frac {1}{r_>}\rho^{\text{WK}}(r_2),
\eeq
with $r_>=\max (r_1,r_2)$ and where appears the spherically averaged charge density of the reference orbital
\beq
\overline{\rho}_{\Omega}^{\text{ref}}(r)=\frac {1}{4\pi}\int d\Omega\;\rho^{\text{ref}}(\vec{x})=\frac{1}{4\pi r^2}\phi^T_{n\kappa_0}(r)\phi_{n\kappa_0}(r).
\eeq

The above atomic energy shift is the quantity that we would like to calculate in a finite Gaussian basis. Before doing so, let us briefly return to the results of Sec.~\ref{sec:defs}. We have established that vacuum polarization can be derived in the Dirac sea picture Eq.~\eqref{eq:DiracVPdensity} from the density matrix $Q$ defined in Eq.~\eqref{eq:dirac_vp_density_matrix}. Now notice that the alternative definition of vacuum polarization in Eq.~\eqref{eq:SchwingerVPdensity} follows in the same manner from :
\beq
	\mathcal{Q} = \frac{P_- - P_+}{2}
\eeq
which as a spectral projector can be computed from a contour integral of the resolvent $(\hat{h}-\zeta)^{-1}$. For the sake of clarity, let us consider the situation of a finite basis, the contours to consider being $\mathcal{C}_-$ and $\mathcal{C}_+$ which encircle the negative and positive parts of the spectrum of $\hat{h}$, respectively. This yields
\beq
	\mathcal{Q} = -\frac{1}{2}\frac{1}{2i\pi}\oint_{\mathcal{C}_--\mathcal{C}_+}\frac{d\zeta}{\hat{h}-\zeta}.
\eeq
The perturbative expansion follows in the exact same manner as 
\beq
	\mathcal{Q} = \sum_{N=0}^{+\infty}\mathcal{Q}^{(N)},
\eeq
where
\beq
	\mathcal{Q}^{(N)} = -\frac{1}{2}\frac{1}{2i\pi}\oint_{\mathcal{C}_--\mathcal{C}_+}\frac{1}{\hat{h}_0-\zeta}\left(-V\frac{1}{\hat{h}_0-\zeta}\right)^N d\zeta
\eeq
and contrary to Eq.~\eqref{eq:vp_density_mat_expansion}, the term $\mathcal{Q}^{(0)}$ does not vanish. \\
We find that 
\beq
	\mathcal{Q}^{(0)} = \frac{P^0_- -P^0_+}{2}
\eeq
and using the completeness relation 
\beq
	P^0_- +P^0_+ = \text{Id},
\eeq
we find that 
\beq
	\mathcal{Q}^{(0)} = P^0_- -\frac{\text{Id}}{2}.
\eeq
The same completeness relation goes for the Coulomb case, where we can write $P_-+P_+=\text{Id}$, and so
\beq
	\mathcal{Q} = P_--\frac{\text{Id}}{2}.
\eeq
This shows that 
\beq\label{eq:Qconn}
\mathcal{Q}-\mathcal{Q}^{(0)}=Q,
\eeq 
and the inspection of the perturbation expansion ensures us that $\mathcal{Q}^{(N)} = Q^{(N)}$ for any $N\geq1$. Therefore, the two definitions of the vacuum polarization density only differ by a constant term, independent of the potential, which is however not trace-class. It is interesting to note that charge-conjugation symmetry was not used when providing the connection Eq.~\eqref{eq:Qconn}.

\section{Basis set considerations}\label{sec:basis}

\subsection{The radial Dirac equation in finite basis}

In the previous section, we have shown how vacuum polarization densities and energy shifts can be computed solely from the eigenstates of the Dirac Hamiltonian. Those are well-known analytically in the case of the Coulomb central field \cite{Mohr1998}. However, generalizing to arbitrary central potentials, and ultimately to molecular ones, will require the use of a numerical approach. For the moment, though, we focus on hydrogen-like ions and restrict our study to the case of radial potentials. What we need is then a set of basis functions with the criteria that they allow to model the physical solutions efficiently and accurately, without, for instance, introducing spurious states\cite{Lewin2014}. \\

The atomic (or, more generally, the central-field) Dirac problem employs the \emph{ansatz} (e.g. \cite[sec.3]{Grant2007})
\beq \label{eq:radial_ansatz}
	\psi_{n\kappa m}(\vec{x})=\frac{1}{r}\begin{bmatrix}
                P_{n\kappa}(r)\xi_{\kappa m}(\Omega) \\
                iQ_{n\kappa}(r)\xi_{-\kappa m}(\Omega)
        \end{bmatrix},
\eeq
with $n$, $\kappa$ and $m$ being respectively the principal, relativistic angular momentum and magnetic quantum numbers, $\xi_{\kappa m}$ is the two-component spherical spinor and $\vec{x}=(r,\Omega)$, where $\Omega=(\theta,\varphi)$ contains the angular spherical coordinates. \\
The angular components being known, the Dirac equation, Eq.~\eqref{eq:Dirac_eq}, then provides the following equation for the radial components 
\beq \label{eq:radial_Dirac_eq}
		\hat{h}_\kappa \phi_{n\kappa} = \epsilon_{n\kappa}\phi_{n\kappa},
\eeq
where
\beq
	\hat{h}_\kappa = \begin{bmatrix}
		mc^2 + V(r) & -c\hbar\left[\dv{r}-\frac{\kappa}{r}\right] \\[3pt]
        c\hbar\left[\dv{r}+\frac{\kappa}{r}\right] & -mc^2 + V(r)
	\end{bmatrix}
\eeq
is the radial Dirac Hamiltonian and 
\beq
	\phi_{n\kappa}(r) = \begin{bmatrix}
                P_{n\kappa}(r) \\
                Q_{n\kappa}(r)
        \end{bmatrix}
\eeq
is the radial Dirac spinor.

We now want to solve the radial Dirac equation, Eq.~\eqref{eq:radial_Dirac_eq}, in a finite basis. We introduce a finite set $\{\chi_{\kappa;\mu}(r)\}_{\mu=1}^N$ of suitable two-component basis functions and project the radial spinor $\phi_{n\kappa}(r)$ onto the subspace spanned by this set
\beq \label{eq:finite_basis}
	\phi_{n\kappa}(r)=\sum_\mu \chi_{\kappa,\mu}(r)c_{\kappa,\mu n}.
\eeq
We endow this subspace with the inner product
\beq
	(f|g) = \int_0^{+\infty} f(r)^\dagger g(r)\mathrm{d} r\qcomma f,g\in L^2(\R_+,\R^2).
\eeq
In this finite basis, the radial Dirac equation, Eq.~\eqref{eq:radial_Dirac_eq},  takes the form of a generalized eigenvalue problem \cite{SalmanThesis}
\beq \label{eq:generalized_eigval_pb}
	H_{\kappa}C_{\kappa}=S_{\kappa}C_{\kappa}\epsilon_{\kappa},
\eeq
where $H_{\kappa},S_{\kappa},C_{\kappa}\in\R^{N\times N}$ are respectively the Hamiltonian, overlap and coefficient matrices, and $\epsilon_{\kappa}\in\text{diag}(\R^N)$ is the diagonal matrix of generalized eigenvalues of $H_{\kappa}$. The overlap and hamiltonian matrix elements are defined as $(S_\kappa)_{\mu\nu} = (\chi_{\kappa,\mu}|\chi_{\kappa,\nu})$ and $(H_\kappa)_{\mu\nu} = (\chi_{\kappa,\mu}|\hat{h}_\kappa|\chi_{\kappa,\nu})$. 
Solving this equation numerically provides us with the coefficients of the Dirac Hamiltonian eigenstates in the finite basis, which we can then use to compute vacuum polarization.

\subsection{Vacuum polarization in finite basis}

With the radial \emph{ansatz}, Eq.~\eqref{eq:radial_ansatz}, in effect, the vacuum polarization density becomes purely radial. Following Eq.~\eqref{eq:vp_finite_diff}, this is also the case for each term in its many-potential expansion. The sum over all states therefore provides a partial-wave expansion, an expansion in $\kappa$, of the vacuum polarization and to each of its orders in $Z\alpha$ \cite{Mohr1998, SalmanThesis}. In particular, the total VP density is given by
\beq \label{eq:radial_vp_density}
	\begin{split}
		\rho^{\text{VP}}(\vec{x}) &= \sum_{\kappa\neq0}\rho^{\text{VP}}_\kappa(r) \\
		\rho^{\text{VP}}_\kappa(r) &= \frac{e\abs{\kappa}}{4\pi r^2}\sum_n \sgn(\epsilon_{n\kappa}-mc^2)\phi_{n\kappa}^\dagger(r)\phi_{n\kappa}(r)
	\end{split}
\eeq
and for the linear part, it follows from the properties of the spherical spinors \cite{SalmanThesis} that
\beq \label{eq:radial_lin_vp_density}
	\begin{split}
		\rho^{\text{VP},(1)}(\vec{x}) &= \sum_{\kappa\neq0}\rho^{\text{VP},(1)}_\kappa(r) \\
		\rho^{\text{VP},(1)}_\kappa(r) &= -\frac{e\abs{\kappa}}{4\pi r^2}\sum_{\substack{\epsilon_{n\kappa}^0 > -mc^2 \\ \epsilon_{\ell\kappa}^0 \le -mc^2}} V_{\kappa,n\ell}^0 \phi_{n\kappa}^{0\dagger}(r)\phi_{\ell\kappa}^0(r),
	\end{split}
\eeq
where we defined
\beq
	V_{\kappa,n\ell}^0 = 4 \frac{(\phi_{n\kappa}^0|V|\phi_{\ell\kappa}^0)}{\epsilon_{n\kappa}^0 - \epsilon_{\ell\kappa}^0}.
\eeq
Finally, the energy shift Eq.~\eqref{eq:atomic_VPshift} also admits a partial-wave expansion 
\beq \label{eq:energy_shift_partial_wave_exp}
	\Delta E^{\text{WK}} = \sum_{\abs{\kappa}>0}\Delta E^{\text{WK}}_\kappa.
\eeq

Now using the finite basis expansion Eq.~\eqref{eq:finite_basis}, we can cast the radial densities Eqs.~\eqref{eq:radial_vp_density} and \eqref{eq:radial_lin_vp_density} as
\beq
	\begin{split} \label{eq:vp_density_matrix_def}
		&r^2\rho^{\text{VP}}_\kappa(r) = -e\Tr\left[D^{\text{VP}}_\kappa\cdot\Omega_\kappa(r)\right] \\
		&D^{\text{VP}}_\kappa = -\frac{\abs{\kappa}}{4\pi}\sum_n \sgn(\epsilon_{n\kappa}-mc^2) c_{\kappa,n}c_{\kappa,n}^\dagger
	\end{split}
\eeq
and
\beq
	\begin{split} \label{eq:lin_vp_density_matrix_def}
		&r^2\rho^{\text{VP},(1)}_\kappa(r) = -e\Tr\left[D^{\text{VP},(1)}_\kappa\cdot\Omega_\kappa(r)\right] \\
		&D^{\text{VP},(1)}_\kappa = \frac{\abs{\kappa}}{4\pi}\sum_{\substack{\epsilon_{n\kappa}^0 > -mc^2 \\ \epsilon_{\ell\kappa}^0 \le -mc^2}}V_{\kappa,n\ell}^0 c_{\kappa,\ell}^0 c_{\kappa,n}^{0\dagger}
	\end{split},
        \eeq
        that is, as the trace of the product of a density matrix with the overlap distribution
        \beq\label{eq:overlap_distribution}
        \Omega_{\kappa,\mu\nu}(r) = \chi_{\kappa,\mu}^T(r)\chi_{\kappa,\nu}(r).
        \eeq
        Likewise, for the density associated with the reference orbital we have
        \beq
        	\begin{split} \label{eq:ref_density}
		&4\pi r^2\overline{\rho}_{\Omega}^{\text{ref}}(r) = -e\Tr\left[D_{\kappa_0}^{\text{ref}}\cdot\tilde{\Omega}_{\kappa_0}(r)\right] \\
		&D_{\kappa_0}^{\text{ref}}= c_{\kappa_0}^{\text{ref}}c_{\kappa_0}^{\text{ref}\dagger}.
	\end{split}
        \eeq
Starting from Eq.~\eqref{eq:atomic_VPshift}, individual $\kappa$ contributions to the Wichmann--Kroll energy shift is given by
\beq \label{eq:energy_shift_dmat}
	\Delta E_\kappa^{\text{WK}} =\frac{1}{4\pi\varepsilon_0} \sum_{\mu\nu\rho\sigma}D^{\text{ref}}_{\kappa_0,\mu\nu}\;\Xi_{\kappa,\nu\mu\sigma\rho}D^{\text{WK}}_{\kappa,\rho\sigma},
\eeq
where appears two-electron integrals
\beq	
	  \Xi_{\kappa,\nu\mu\sigma\rho} = 4\pi\int_0^{+\infty}d r_1\int_0^{+\infty}d r_2\;
          \tilde{\Omega}_{\kappa_0,\nu\mu}(r_1)\frac{1}{r_>}\Omega_{\kappa,\sigma\rho}(r_2)
          \eeq
and the density matrix of the reference orbital          
\beq
D_{\kappa_0}^{\text{ref}}=c_{\kappa_0}^{\text{ref}}c_{\kappa_0}^{\text{ref}\dagger}.
\eeq
It may be noted that we have placed a tilde over the overlap distribution $\tilde{\Omega}_{\kappa_0}$, associated with the reference orbital, since we in
practice will use different basis sets for the reference orbital and the VP density, as will be explained in the next section.

\subsection{Basis set construction}

An immediate concern when constructing basis sets for relativistic calculations is the coupling between large and small components.
From Eq.~\eqref{eq:radial_Dirac_eq}, it is seen to be formally energy-dependent
\begin{eqnarray}
	Q_{n\kappa}(r) &= \frac{\hbar}{mc} \left[1+\frac{\epsilon_{n\kappa} - V(r)}{mc^2}\right]^{-1}\left[\dv{r} + \frac{\kappa}{r}\right] P_{n\kappa} \label{eq:PtoQ}\\
	P_{n\kappa}(r) &= \frac{\hbar}{mc} \left[1-\frac{\epsilon_{n\kappa} - V(r)}{mc^2}\right]^{-1}\left[\dv{r} - \frac{\kappa}{r}\right] Q_{n\kappa}. \label{eq:QtoP}
\end{eqnarray}
The usual prescription, known as Restricted Kinetic Balance (RKB)\cite{Schwarz1982,stanton:kinbal,Dyall_CPL1990}, considers the non-relativistic limit of Eq.~\eqref{eq:PtoQ} as a constraint on the small component of the basis set elements. In the case of an extended nucleus, the potential energy $V$ is bounded and assumed to obey $V(r)\ll mc^2$. Setting $\epsilon_{n\kappa}= mc^2 + O(c^0)$, the non-relativistic limit yields an energy-independent expression
\beq
\lim_{c\rightarrow+\infty}c Q_{n\kappa}(r) \simeq \frac{\hbar}{2m}\left[\dv{r}+\frac{\kappa}{r}\right]P_{n\kappa}.
\eeq
At the basis-set level this translates into
\beq
	\phi_{n\kappa}^{RKB}(r) = \sum_{\mu=1}^N\left[c_{\kappa,\mu n}^L\begin{pmatrix}
		\pi_{\kappa\mu}^L(r) \\ 0
	\end{pmatrix} + c_{\kappa,\mu n}^S\begin{pmatrix}
		0 \\ \Tilde{\pi}_{\kappa\mu}^S(r)
	\end{pmatrix}\right],
\eeq
where a small component basis function $\Tilde{\pi}_{\kappa\mu}^S(r)$ is obtained from a large component one $\pi_{\kappa\mu}^L(r)$ as
\beq \label{eq:coupled_small_component}
	\Tilde{\pi}_{\kappa\mu}^S(r) = \Tilde{\mathcal{N}}_{\kappa\mu}^S\left[\dv{r} + \frac{\kappa}{r}\right]\pi_{\kappa\mu}^L(r),
        \eeq
where $\Tilde{\mathcal{N}}_{\kappa\mu}^S$ is the normalization factor. This construction assures that the kinetic energy is correctly reproduced in the non-relativistic limit\cite{dyall:kinbal}; the correct coupling of large and small components at finite speed of light requires some flexibility in the basis\cite{visscher:kinbal}. However, by setting $\epsilon_{n\kappa}= mc^2 + O(c^0)$, this basis set only properly describes positive-energy solutions. If one is interested in the negative part of the spectrum, one may instead start from from Eq.~\eqref{eq:QtoP}, setting $\epsilon_{n\kappa}= -mc^2 + O(c^0)$ and take the non-relativistic limit. In the resulting Inverse Kinetic Balance (IKB) scheme\cite{sun_TCA2011} large-component basis functions $\Tilde{\pi}_{\kappa\mu}^L(r)$ are generated from small-component basis functions $\pi_{\kappa\mu}^S(r)$ according to
\beq \label{eq:coupled_large_component}
	\Tilde{\pi}_{\kappa\mu}^L(r) = \Tilde{\mathcal{N}}_{\kappa\mu}^L\left[\dv{r}-\frac{\kappa}{r}\right]\pi_{\kappa\mu}^S(r).
\eeq
The Dual Kinetic Balance (DKB) scheme proposed by Shabaev \textit{et al.} \cite{Shabaev2004} treats positive and negative energies on an equal footing by expanding radial functions as
\beq
	\phi_{n\kappa}^{DKB}(r) = \sum_{\mu=1}^N\left[c_{\kappa,\mu n}^L\begin{pmatrix}
		\pi_{\kappa\mu}^L(r) \\ \Tilde{\pi}_{\kappa\mu}^S(r)
	\end{pmatrix} + c_{\kappa,\mu n}^S\begin{pmatrix}
		\Tilde{\pi}_{\kappa\mu}^L(r) \\ \pi_{\kappa\mu}^S(r)
	\end{pmatrix}\right].
 \eeq
 Three other schemes are worth mentioning: i) Recently, Grant and Quiney proposed a DKB-like scheme which introduces an energy-dependence in the basis\cite{GrantQuiney2022atoms}. ii) Previously, Dyall proposed a scheme, ``dual atomic balance'', where the basis is generated from positive-energy and negative-energy electronic solutions obtained with RKB and IKB, respectively\cite{Dyall_CP2012}. iii) Our alternative expression for the VP-density, Eq.~\eqref{eq:VP-RKB}, suggests a third scheme, namely using electronic and positronic positive-energy solutions, both generated using RKB. This has the conceptual advantage of only referring to observable solutions of the Dirac Hamiltonian.
 
For the scalar basis functions $\pi_{\kappa\mu}^L$ and $\pi_{\kappa\mu}^S$, we adopt normalized Gaussian functions, 
	\begin{eqnarray}
		\pi_{\kappa\mu}^{L}(r)&=\mathcal{N}_{\kappa\mu}^{L}r^{\ell_{\kappa}+1}e^{-\zeta_{\mu}r^{2}}\\
		\pi_{\kappa\mu}^{S}(r)&=\mathcal{N}_{\kappa\mu}^{S}r^{\ell_{-\kappa}+1}e^{-\zeta_{\mu}r^{2}},
	\end{eqnarray}
        where $\ell_{\kappa}=|\kappa|+\frac{1}{2}(\mbox{sgn}(\kappa)-1)$ \cite{Biedenharn_PhysRev.126.845}. Gaussian functions, introduced by Boys in 1950\cite{Boys_RSPA1950}, are perfect for our problem. First, they are very easy to handle since every matrix element can be computed analytically. Furthermore, in the context of QED, it is convenient that the Fourier transform of a Gaussian is a Gaussian. The power of $r$ is chosen so that the basis reproduces the correct small-$r$ asymptotic behavior of the exact Dirac solution for an extended nucleus \cite{Ishikawa1987}. In this work Ishikawa and Quiney showed that the leading terms of the series expansion of the exact solution coincide with those of a Gaussian function within the nuclear region. This suggests Gaussians as a very natural choice of basis function for the description of the wavefunction in the nuclear region, which is precisely where vacuum polarization is the strongest.

In previous work Salman and Saue showed that the calculation of the VP density according to Eq.~\eqref{SchwingerVP2} requires relativistic basis sets that comply with charge conjugation ($\cal C$) symmetry\cite{Salman_sym2020,Salman_PRA2023}, such that the VP density vanishes in the free-particle case (see also Ref. \citenum{GrantQuiney2022atoms}). One such basis set construction is DKB with the additional requirement that the same list of exponents should be used for $\pm\kappa$ ($j$-basis)\cite{Dyall_TCA1996}, that is, $\pi_{\kappa\mu}^{L}=\pi_{-\kappa\mu}^{S}$. Restricted kinetic balance (RKB), which favors positive-energy solutions, \emph{a priori} fails to provide $\cal C$ symmetry. However, excellent results can be obtained by using a modified expression
\begin{equation}
  \rho^{\text{VP}}_{\cal C}(\vec{x},Z)=\frac{1}{2}\left[\rho^{\text{VP}}(\vec{x},Z)-\rho^{\text{VP}}(\vec{x},-Z)\right],
\end{equation}
that enforces $\cal C$ symmetry. Written out in the notation of Section \ref{sec:defs}, the expression reads
\begin{align}
  \rho^{\text{VP}}_{\cal C}(\vec{x},Z)=&-\frac{e}{4}\left[\left\{n^{(-)}_e(\vec{x},Z)-n^{(+)}_e(\vec{x},Z)\right\}\right.\notag\\
    &-\left.\left\{n^{(-)}_e(\vec{x},-Z)-n^{(+)}_e(\vec{x},-Z)\right\}\right].
\end{align}
We do not expect $n^{(-)}_e(\vec{x},\pm Z)$ to be well described in RKB basis, but apparently errors cancel out. A simpler alternative would be to use Eq.~\eqref{eq:VP-RKB}. In the present work, however, we shall use DKB.

It remains to choose the list of exponents to be used in our calculations. The most compact basis sets are obtained by generation of exponents through energy optimization (see, for instance, Ref.~\cite{Jensen_2007ICC}).
However, such a procedure optimizes the description of \textit{occupied} orbitals in an atom or a molecule, but is clearly less suitable for the sum over states appearing in the expressions for the VP-density seen above.
In the present work, for the calculation of the WK energy shift, Eq.~\eqref{eq:atomic_VPshift}, we have therefore chosen to expand the reference orbital $\phi_{n\kappa_0}$ in an energy-optimized basis, whereas for
the orbital generating the VP density we use an even-tempered basis\cite{RuedenbergRaffenetti}, where exponents are in a geometric progression
\beq \label{eq:even_temp_exp}
	\zeta_\mu = \zeta_{\min}\left(\frac{\zeta_{\max}}{\zeta_{\min}}\right)^{\frac{\mu-1}{N-1}} = \zeta_{\min}\beta^{\mu-1}\qcomma \mu=1,\dots,N.
\eeq
Here $\beta$ is the geometric ratio that dictates the density of the basis. This choice is very practical since the basis is only defined by three parameters: the basis size $N$, $\zeta_{\min}$ and either $\beta$ or $\zeta_{\max}$. Further details on the generation of the VP basis will be given in the following.

\subsection{Numerical analysis} \label{sec:numerical_noise}

We have established in Eqs.~\eqref{eq:vp_density_matrix_def}, \eqref{eq:lin_vp_density_matrix_def} and \eqref{eq:ref_density} that the densities we consider can all be computed as the trace of the matrix product between a density matrix $D$ and an overlap distribution matrix $\Omega(r)$
\beq
	r^2\rho(r) = \Tr(D\cdot\Omega(r)),
        \eeq
        with a density matrix on the generic form
        \beq\label{eq:general_density_matrix}
        D = \sum_{\ell n}\lambda_{\ell n}c_{\ell}c_{n}^\dagger,
        \eeq
where the $c_n$ are the vectors of coefficients of the radial Dirac wavefunction in finite basis, and the $\lambda_{\ell n}$ are real coefficients. The coefficients of a wavefunction in a non-orthonormal basis cannot be normalized in the sense of the Euclidean norm for them to describe a physical state. We will now describe the consequences of linear dependencies on their norm, and on the occurrence of numerical noise in our calculations. \\

The coefficients $c_n$ are solutions to the generalized eigenvalue problem in Eq.~\eqref{eq:generalized_eigval_pb}, which can be solved by performing an orthonormalization of the basis. 
Let us consider a basis set $\{\chi_\mu(r)\}_{\mu=1}^N$, in which the radial Dirac equation takes the form of Eq.~\eqref{eq:generalized_eigval_pb}. We seek an invertible matrix $V\in GL(N,\R)$ such that
\beq \label{eq:orthonormalization}
	V^\dagger S V = I_N.
\eeq
Löwdin showed that the general form for such a matrix is \cite{Lowdin_Ortho}
\beq \label{eq:Lowdin_ortho_matrix}
	V = S^{-1/2}U,
\eeq
where $U\in O(N)$ is an arbitrary orthogonal matrix. The most common choices for $V$ are the symmetric, canonical, and Cholesky orthonormalizations. The symmetric orthonormalization $V_{\text{sym}} = S^{-1/2}$ was shown by Carlson and Keller \cite{Carlson} to satisfy a least-square condition, the minimization of the Frobenius norm $\norm{I_N - V}_F$. This guarantees the orthonormalized basis to be as close as possible to the original one. Next, by diagonalizing the overlap matrix, $S = W s W^\dagger$, with $W\in O(N)$ and $s\in\text{diag}(R_+^N)$, one can define the canonical orthonormalization $V_{\text{can}} = S^{-1/2}W = Ws^{-1/2}$. This choice is particularly convenient for dealing with linear dependencies, as $s$ and $V$ can be made rectangular by removing the columns of $s^{-1/2}$ falling below a pre-selected threshold. In our calculations, however, we did not follow this procedure, as it proved to be very unstable for the calculations of VP densities. Lastly, the Cholesky orthonormalization $V_{\text{Chol}}=L^{-\dagger}$, where $L$ is the lower-triangular matrix stemming from the Cholesky decomposition $S=LL^\dagger$, is often preferred for its numerical stability \cite[sec.4.2.7.]{Golub}. It is in essence equivalent to the Gram-Schmidt procedure, which can be seen by taking the QR decomposition of $V$ in Eq.~\eqref{eq:orthonormalization}.

\begin{figure}[h]
\centering
\includegraphics[width=1.\linewidth]{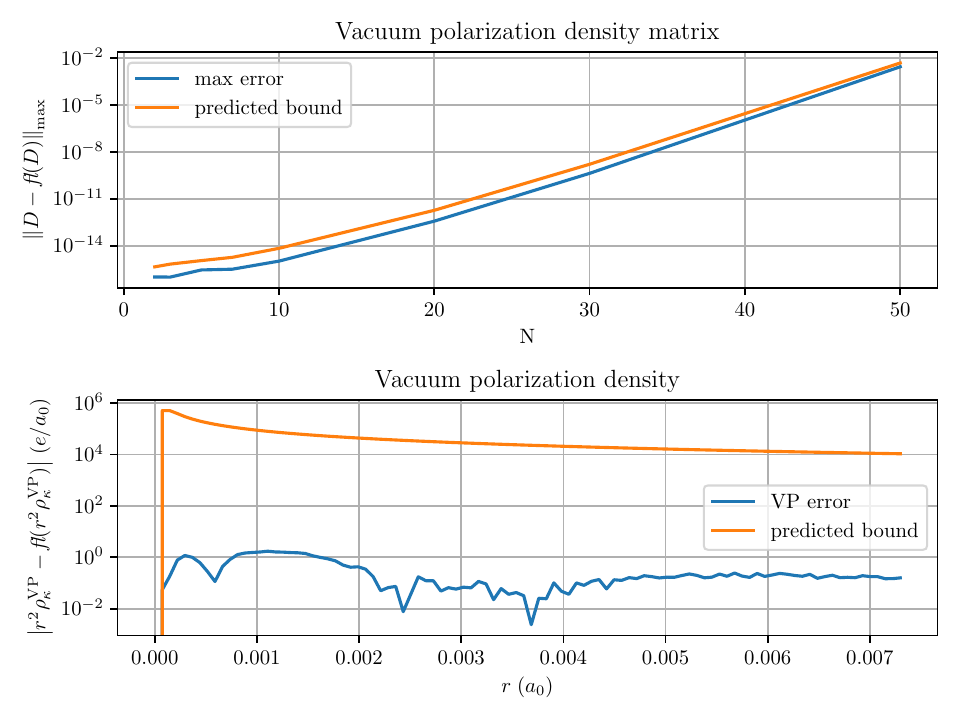}
\caption{Numerical evaluation of the predicted error bounds, Eqs.~\eqref{eq:dmat_err_bound} and \eqref{eq:density_err_bound}, for the vacuum polarization density matrix and density. Evaluation in double-precision for $Z=92$, $\kappa=-1$ and a point nucleus for different basis sizes. The bases are even-tempered with $\zeta_{\min}=10^3\,a_0^{-2}$ and $\zeta_{\max}=10^8\,a_0^{-2}$. 
Errors are computed by comparing the double-precision calculation to an arbitrary-precision floating-point arithmetic calculation with 50 relevant digits. 
The vacuum polarization density in the bottom plot is computed with $N=50$ exponents ($\beta=1.26$).}
\label{fig:numerical_bounds_vp}
\end{figure}

Now transforming to our orthonormal basis, the transformed problem reads
\beq
	H^\prime C^\prime = C^\prime \epsilon,
\eeq
where
\beq \label{eq:transformed_quantities}
	H^\prime = V^\dagger H V\qcomma C = V C^\prime.
\eeq
In this orthonormal basis, Eq.~\eqref{eq:generalized_eigval_pb} yields a regular eigenvalue problem, solvable via standard algorithms, whose solution is an orthogonal matrix $C^\prime\in O(n)$. One then transforms back into the original basis by using Eq.~\eqref{eq:transformed_quantities}. For any matrix norm that is unitarily invariant we then have
\beq
	\norm{C} = \norm{V} = ||S^{-1/2}||,
\eeq
where the last identity follows from Eq.~\eqref{eq:Lowdin_ortho_matrix}. Therefore, in the case of the $2$-norm,
\beq \label{eq:norm_coeffs_eigvals}
	\norm{C}_2 = \norm{V}_2 = \frac{1}{\sqrt{\min \sigma(S)}},
\eeq
where $\sigma(S)$ is the spectrum of $S$. We see that the norm of $C$ will diverge as $\min\sigma(S)$ goes to $0$, which is the case when two or more basis functions become linearly dependent. 
This is the main source of numerical noise in our finite basis calculations. In floating-point arithmetic, floating-point numbers are represented on $b$ bits with $u=2^{-b}$ being the unit roundoff. Let us denote by $fl(\cdot)$ the floating-point representation of any given operation. Assuming $uN\ll1$, we can perform a rounding error analysis \cite{Blanchard, Higham} to show that 
\beq \label{eq:dmat_err_bound}
	\norm{D - fl(D)}_{\max} \leq \frac{uN}{\min\sigma(S)} + \mathcal{O}(u^2).
\eeq
For the density $r^2\rho^{\text{VP}}$, the bound obtained in this fashion 
\beq \label{eq:density_err_bound}
	\abs{r^2\rho(r) - fl(r^2\rho(r))} \leq uN \norm{D}_{2,\infty}\norm{\Omega(r)}_{2,1} + \mathcal{O}(u^2),
\eeq
is too loose, as it is designed to capture the worst-case scenario and hence does not account for error cancellations at play in the density computations. The two bounds of Eqs.~\eqref{eq:dmat_err_bound} and \eqref{eq:density_err_bound} are reported on Fig.~\ref{fig:numerical_bounds_vp}. Still, the latter is in and by itself not completely useless. In the words of Higham \cite[sec.3.2]{Higham}, ``\textit{The purpose [of rounding error analysis] is to show the existence of an a priori bound for some appropriate measure of the effects of rounding errors on an algorithm. Whether a bound exists is the most important question. Ideally, the bound is small for all choices of problem data. If not, it should reveal features of the algorithm that characterize any potential instability and thereby suggest how the instability can be cured or avoided}''. \\
The lesson here is that the precision on vacuum polarization is inversely proportional to the smallest eigenvalue of the overlap matrix $S$. This is the decisive factor for our choice of basis. In principle, we would like to consider bases for the vacuum as dense as possible, but this will cause the smallest eigenvalue of $S$ to fall to $0$, and hence loss of numerical precision. For this reason, we should also consider using different bases for the vacuum and the reference orbital in energy shift calculations, as their two characteristic radii are very different, the former on the order of a reduced Compton wavelength ($\lambdabar = \hbar/mc $). Furthermore, the reference state does not need bases as dense as the vacuum polarization density. We thereby gain both in  computation efficiency and precision. In what follows, we will present a quantitative method to relate the density of our bases with this numerical noise.

\subsection{Linear dependencies in the even-tempered basis sets}\label{sec:lindep}

We now have an idea of the relation between the smallest eigenvalue of the overlap matrix and the magnitude of floating-point errors. This provides a formal bound to how dense we can select our basis functions. But in order to provide a quantitative criterion, we still need to investigate the magnitude of this smallest eigenvalue for the basis schemes considered. 
As stated earlier, we chose to model the vacuum using normalized even-tempered basis sets, which are defined by only three parameters, which simplifies the analysis.
Let us then start by considering the simpler case of the non-relativistic hydrogen-like atom. 

\subsubsection{Non-relativistic hydrogen atom}

Solving the Schrödinger equation for a non-relativistic hydrogen-like ion in a finite Gaussian basis
\beq
	\pi_{\ell}(r,\zeta)	= \mathcal{N}_{\ell\zeta}r^{\ell+1}e^{-\zeta r^2}
\eeq
yields the following generalized eigenvalue problem :
\beq
	H_\ell C_\ell = S_\ell C_\ell\epsilon_\ell.
\eeq
with the overlap matrix elements given by
\beq \label{eq:Schrodinger_overlap}
	(S_\ell)_{\mu\nu} = \left(\frac{2\sqrt{\zeta_\mu\zeta_\nu}}{\zeta_\mu+\zeta_\nu}\right)^{\ell+\frac{3}{2}}.
\eeq	

        It was noted by Reeves that the overlap of normalized Gaussians only depends on the exponent ratio, which thereby suggested selection of exponents according to geometric progression\cite{Reeves_JCP1963a,Reeves_JCP1963b} (see Eq.~\eqref{eq:even_temp_exp}). This basis set construction was formalized under the (musical) name of even-tempered basis by Ruedenberg and co-workers\cite{RuedenbergRaffenetti}. Writing the radial function as an integral transform
\beq \label{eq:gcm1}
   u_{n\ell}(r)=\int_0^{\infty}{\,}d\zeta\pi_{\ell}(r,\zeta)f_{n\ell}(\zeta),
\eeq
and introducing a change of variable
\beq \label{eq:gcm2}
   u_{n\ell}(r)=\int_{-\infty}^{\infty}{\,}d(\ln\zeta)\pi_{\ell}(r,\zeta)\tilde{f}_{n\ell}(\zeta),
     \eeq
     the same authors point out that an even-tempered basis $\left\{\zeta_\mu=\zeta_{\min}\beta^{\mu-1}\right\}_{\mu=1}^N$ is automatically generated upon discretization on an even-spaced grid. This observation connects to the generator coordinate method, originally formulated in nuclear physics\cite{Hill_PhysRev.89.1102,Griffin_PhysRev.108.311}, where weight functions $f(\zeta)$ are variational parameters. Upon adaptation to atomic and molecular calculations, it was noted that the original form, Eq.~\eqref{eq:gcm1}, worked well with Slater exponential functions, but not with Gaussian-type ones, for which the modified transform Eq.~\eqref{eq:gcm2} was introduced in order to obtain satisfactory convergence\cite{Mohallem_IJQC1986,Mohallem_ZfPD1986}. Ruedenberg and co-workers\cite{RuedenbergRaffenetti} also point out that their observation suggests that the complete basis set limit is obtained as
     \beq\label{eq:ebas_lim}
     \zeta_{\min}\rightarrow 0^+,\beta\rightarrow 1^+, N\rightarrow\infty .
     \eeq
A more formal proof has been given by Klahn\cite{Klahn_JCP1985}.
     
\begin{figure}[h]
\centering
\includegraphics[width=1.\linewidth]{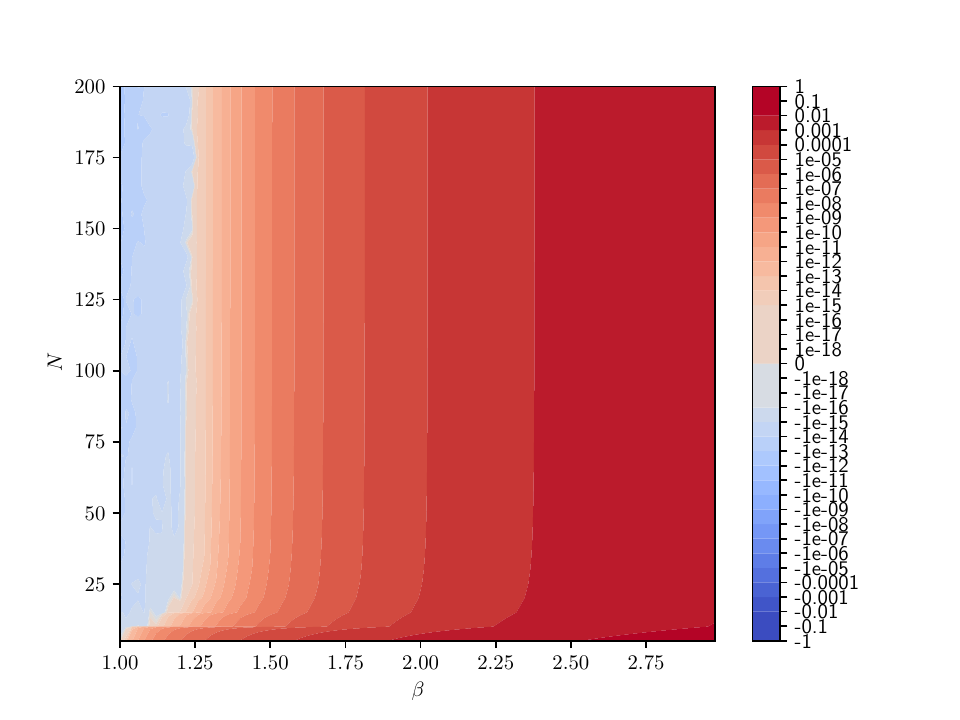}
\caption{Heatmap representing the order of magnitude of the smallest eigenvalue of $S$ for $\ell=0$ as a function of $\beta$ and the basis size $N$.}
\label{fig:non_relativistic_heatmap}
\end{figure}

In a normalized even-tempered basis set, the overlap matrix indeed takes the simpler form
\beq
	(S_\ell)_{\mu\nu} = \left(\frac{2\beta^{\frac{\mu-\nu}{2}}}{1+\beta^{\mu-\nu}}\right)^{\ell+\frac{3}{2}},
\eeq
forming a Toeplitz matrix only dependent on $\beta$ and $N$ (see also Ref.~\cite[Eq.~(8.2.15)]{Helgaker}).
This simplifies the description of the relationship between the basis and the smallest eigenvalue of
the overlap matrix $S_\ell$. In Fig.~\ref{fig:non_relativistic_heatmap} this relation is depicted as a heatmap for $\ell=0$.

\begin{figure}[h]
\centering
\includegraphics[width=1.\linewidth]{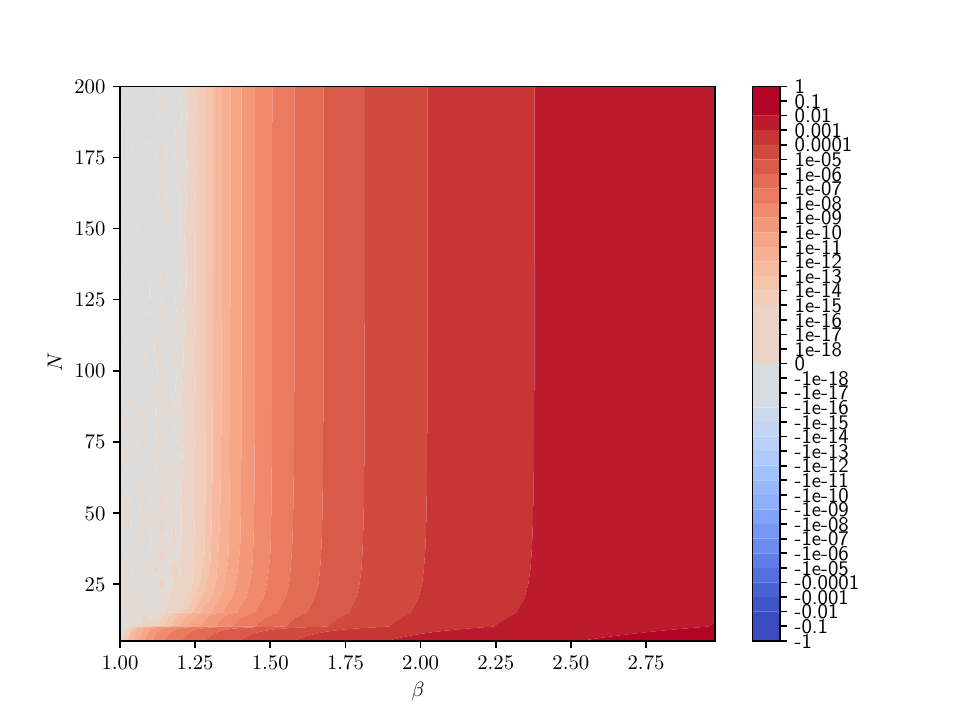}
\caption{Same calculation as in Fig.~\ref{fig:non_relativistic_heatmap}, but with the diagonalization performed using the SVD algorithm.}
\label{fig:non_relativistic_heatmap_svd}
\end{figure}

The data from Fig.~\ref{fig:non_relativistic_heatmap} was obtained with FORTRAN in double precision using the LAPACK routine \textit{dsyev}\cite{LAPACK99}. We can first observe that with very small $\beta$, typically $\beta < 1.25$, negative eigenvalues appear in the spectrum of $S$, which is a sign of numerical instability since $S$ is a Gram matrix and is therefore positive semi-definite. We can avoid this defect by instead using singular value decomposition (SVD; LAPACK routine \textit{dgesvd}) for diagonalization, as $S$ is also hermitian, see Fig.~\ref{fig:non_relativistic_heatmap_svd}. The negative singular values are all set to zero by the algorithm, which allows the calculation of $S^{-1/2}$. 

\begin{figure}[h]
\centering
\includegraphics[width=1.\linewidth]{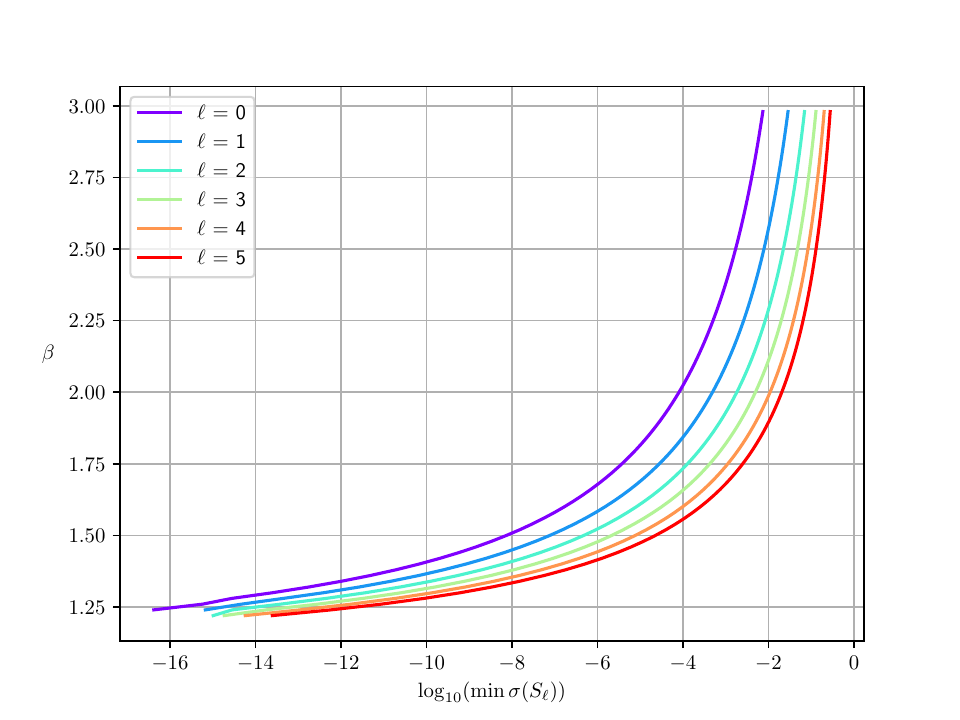}
\caption{Relation between the smallest eigenvalue of the overlap matrix and $\beta$ in an even-tempered Gaussian basis for different values of $l$.}
\label{fig:beta_select}
\end{figure}

For large enough bases, see Fig.~\ref{fig:non_relativistic_heatmap}, we observe vertical lines, a sign that the magnitude of the smallest eigenvalue of $S$ becomes roughly independent of the basis size $N$ and depends primarily on $\beta$.
This then allows us to draw the simpler plot shown in Fig.~\ref{fig:beta_select}. It is interesting to notice that for a fixed $\beta$, $\min\sigma(S)$ increases with $\ell$. Therefore, our basis should be selected only for the most critical case $\ell=0$.

\subsubsection{Restricted Kinetic Balance}

Now, considering the solutions to the radial Dirac equation, we can start with the simpler case of the Restricted Kinetic Balance (RKB) basis scheme. The large and small component blocks of the overlap matrix for normalized Gaussian basis elements can be obtained, with respect to Eq.~\eqref{eq:Schrodinger_overlap}, as
\beq
	S_\kappa^{LL} = S_{\ell_\kappa}\qcomma S_\kappa^{SS} = S_{\ell_{\kappa}+1}.
\eeq
The latter result is obtained using
\beq
(\kappa+\ell_\kappa+1)(\kappa-\ell_\kappa)=0.        
\eeq
The error analysis can therefore be performed using the non-relativistic heatmap. \\

\begin{figure}[h]
\centering
\includegraphics[width=1.\linewidth]{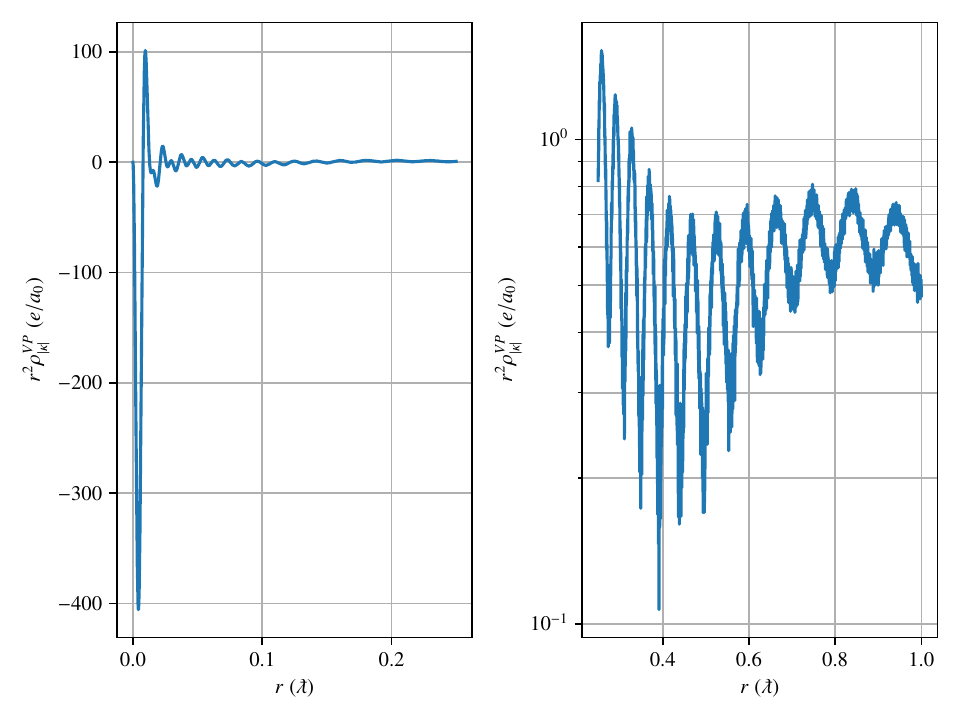}
\caption{Vacuum polarization density in \ce{U^{91+}} with a point nucleus and $\abs{\kappa}=1$. Computation performed in DKB with the basis $\zeta_{\min}=10^3\,a_0^{-2}$, $\zeta_{\max}=10^8\,a_0^{-2}$ and $N=50$ ($\beta \simeq 1.26$). We represent the short and long range behavior, separated around $r=\lambdabar/4$. The oscillations at play are not only a feature of the finite basis representation, but also the consequence of the presence of the non-regularized linear contribution in $r^2\rho_{\abs{\kappa}}^{\text{VP}}$. The point here is to display the typical magnitude of the VP density, and the arising of numerical noise, here in the long range regime.}
\label{fig:vp_noise}
\end{figure}

\begin{figure}[h]
\centering
\includegraphics[width=1.\linewidth]{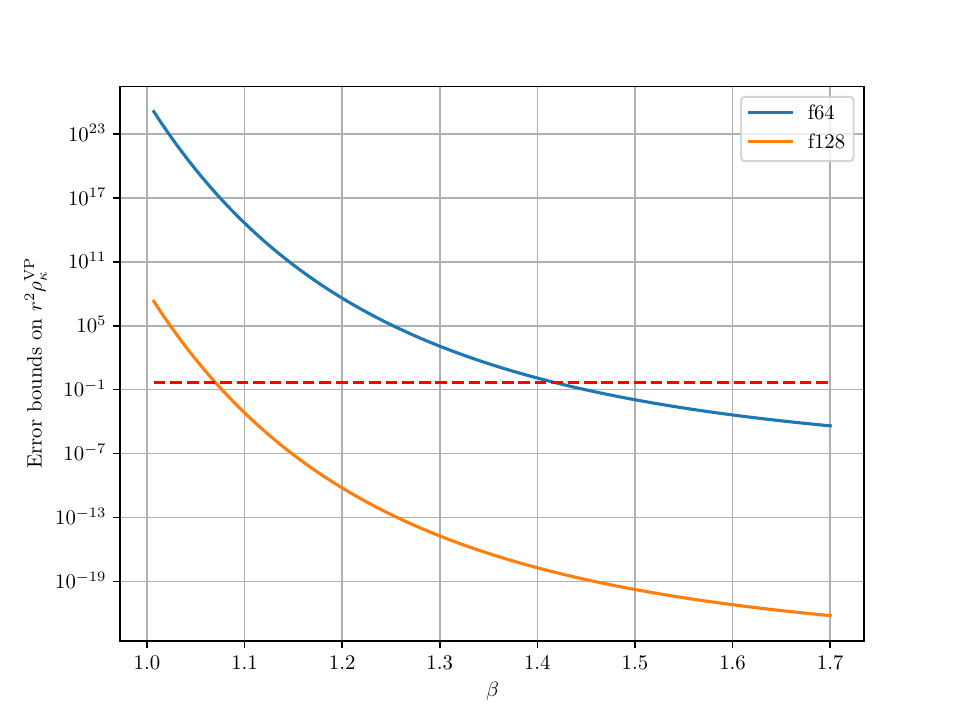}
\caption{Bound of Eq.~\eqref{eq:density_err_bound} evaluated for \ce{U^{91+}} at $r = \lambdabar$, with $\zeta_{\min} = 10^3\,a_0^{-2}$, plotted for double (f64) and quadruple (f128) precision against $\beta$, and compared to the typical magnitude of the total VP density of $0.5\,e/{a_0}$ at this range (red dashed line), as seen from Fig.~\ref{fig:vp_noise}. This provides a good estimation for the density of the basis at which floating-point calculations of the VP density start to suffer from numerical noise.}
\label{fig:noise_estimation}
\end{figure}

In addition, we computed an exact bound for $\norm{\Omega_{\kappa}^{\text{RKB}}(r)}_{2,1}$ in Eq.~\eqref{eq:density_err_bound}, which yields
\beq \label{eq:NR_VP_bound}
	\begin{split}
	&\abs{r^2\rho(r) - fl(r^2\rho(r))} \leq \\
	&\frac{uN\zeta_{\min}}{\min\sigma(S)}C(\kappa)\sqrt{\frac{1-\beta^{2N}}{1-\beta}}\frac{1-\beta^N}{1-\beta^{1/2}} + \mathcal{O}(u^2), \\
	&C(\kappa) = \frac{2^{3/2}\ell_\kappa^{\ell_\kappa}e^{-\ell_\kappa}}{\Gamma(\ell_\kappa+\frac{1}{2})}.
	\end{split}
\eeq
This is the best analytical bound that we could find on $\norm{\Omega_\kappa^{RKB}(r)}_{2,1}$, but it is still too large for any practical estimation of the magnitude of numerical noise. However, if we simply plot the bound in Eq.~\eqref{eq:density_err_bound} while computing numerically the $2,1$-norm, and compare it to the expected magnitude of $0.1\,e/{a_0^3}$ for the VP density at long range (cf. Fig.~\ref{fig:vp_noise}), we see on Fig.~\ref{fig:noise_estimation} that double-precision calculations should start to suffer from numerical noise around $\beta=1.4$, whereas quadruple-precision calculations become unreliable around $\beta=1.1$. These estimations have been confirmed by our energy shift calculations, as seen in Fig.~\ref{fig:wk_eshift_Rn}.

\subsubsection{Dual Kinetic Balance}

\begin{figure}[h]
\centering
\includegraphics[width=1.\linewidth]{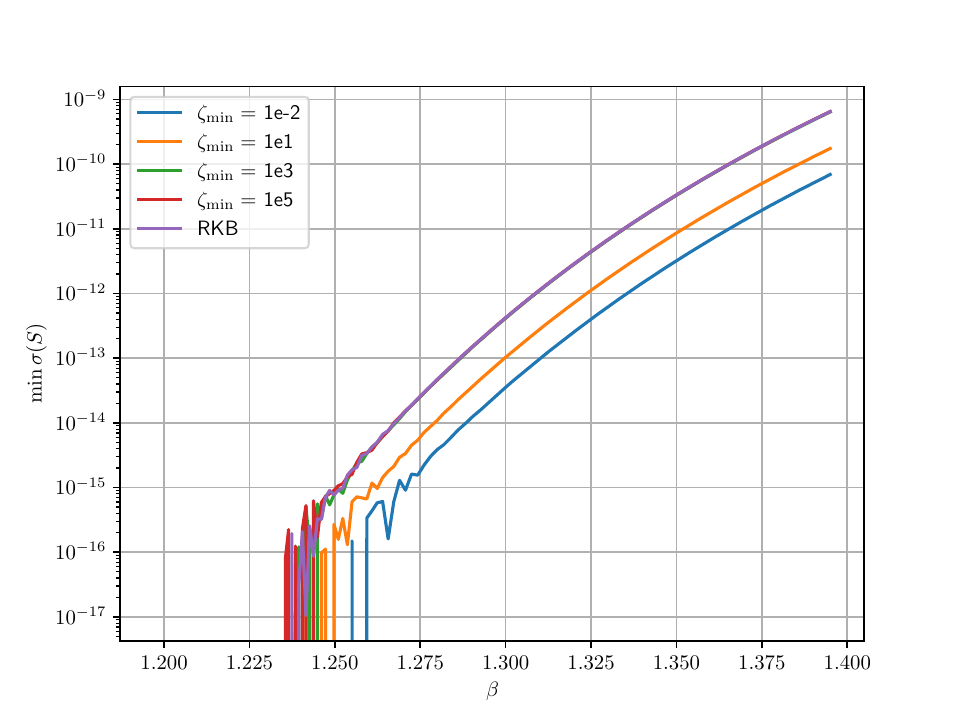}
\caption{Lowest eigenvalue of the DKB overlap matrix for $\kappa=1$, a basis of size $N=100$ and different values of $\zeta_{\min}$. Calculations performed in double precision.}
\label{fig:DKB_lowest}
\end{figure}

The case of Dual Kinetic Balance is harder to analyze. In this case, the overlap matrix elements now have an explicit dependence on $\zeta_{\min}$ and $\zeta_{\max}$. But as it is the case for RKB, we can assume that the dependence on the basis size $N$ vanishes for large enough bases, a consequence of the vertical stripes in Fig.~\ref{fig:non_relativistic_heatmap}. We can then plot the $\beta$ dependence of the lowest eigenvalue of the overlap matrix $S$ for different values of $\zeta_{\min}$, see Fig.~\ref{fig:DKB_lowest}.

We see that the lowest eigenvalue of $S$ increases with $\zeta_{\min}$; small exponents are therefore more prone to numerical instabilities. Interestingly, Fig.~\ref{fig:DKB_lowest} suggests that the smallest eigenvalue of the DKB overlap matrix converges to that of the RKB matrix for large $\zeta_{\min}$, so DKB is always worse-conditioned than RKB. In addition, we see that the double-precision eigensolver becomes unreliable below $\beta=1.275$. This sets a hard limit to the basis density we can require for double-precision calculations. In fact, Fig.~\ref{fig:noise_estimation} suggests that instabilities occur already at $\beta=1.4$. It is clear from Eqs.~\eqref{eq:dmat_err_bound} and \eqref{eq:density_err_bound} that the only possibility then is to decrease $u$, \textit{i.e.} increase the numerical precision. As we have seen above, DKB calculations in quadruple precision are expected to be robust against numerical noise for $\beta$ larger than $1.1$.

\begin{figure}[h]
\centering
\includegraphics[width=1.\linewidth]{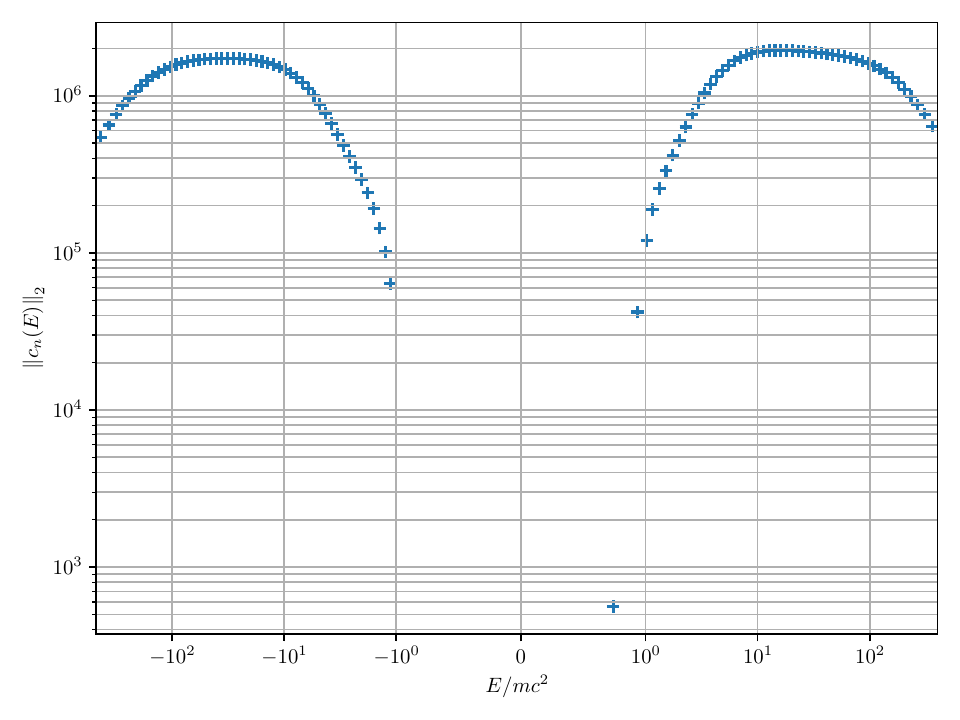}
\caption{Representation of the coefficient 2-norm $\norm{c_n}_2$ of the Dirac Hamiltonian eigenstates as a function of their energy. This computation was performed on \ce{U^{91+}} for $\kappa = -1$ and a shell nucleus with $r_n=5.751\,\text{fm}$ using an even-tempered basis in DKB with $\zeta_{\min}=10^3\,a_0^{-2}$, $\zeta_{\max}=10^8\,a_0^{-2}$ and $N = 50$ ($\beta \simeq 1.26$).}
\label{fig:coeffs_norm_energy}
\end{figure}

In Figure \ref{fig:coeffs_norm_energy} we trace the 2-norm $\norm{c_n}_2$ of the coefficients of eigensolutions as a function of their energy. The calculation was carried out in DKB with $\zeta_{\min}=10^3\,a_0^{-2}$, $\zeta_{\max}=10^8\,a_0^{-2}$ and $\beta \simeq 1.26$, where $\min\sigma(S)$ is expected at $10^{-14}$ according Fig.~\ref{fig:DKB_lowest}. From Eq.~\eqref{eq:norm_coeffs_eigvals} we expect the largest norm amongst the coefficients to reach $10^6$, which is indeed what we see in Fig.~\ref{fig:coeffs_norm_energy}. Interestingly, continuum states suffer from a significantly larger norm than bound states.

\section{Implementation and computational details}\label{sec:comp}

The following calculations are carried out in atomic units, with the physical constants following the CODATA 2022 values \cite{CODATA2022}. In particular, we use the following values for the inverse fine structure constant $\alpha^{-1}=137.035999177$, Bohr radius $a_0=5.29177210903\times10^4\,\text{fm}$ and Hartree energy $E_h=27.211386245981\,\text{eV}$. \\

Following Persson \textit{et al.} \cite{Persson1993}, we consider two different nuclear models : the shell nuclear model which is associated with the density \cite{Salman_PRA2023}
\beq
	\nu^{\text{shell}}(r) = \frac{1}{4\pi r_n^2}\delta(r-r_n),
\eeq
and the uniformly charged sphere with \cite{Visscher1997}
\beq
	\nu^{\text{unif.}}(r) = \frac{3}{4\pi R_0^3}\mathds{1}_{\{r\leq R_0\}}(r),
\eeq
where $r_n$ is the root-mean square of the nuclear distribution, and $R_0 = r_n\sqrt{\frac{5}{3}}$. \\

The code used for these calculations consists of a \textit{Python} interface that calls C routines for the most compute-intensive parts. The user interface is coded as a \textit{Python} library, and allow for arbitrary-precision calculations with the mpmath library \cite{mpmath}. In this interface, the user defines the basis and nuclear model considered, and the radial Dirac equation is solved in arbitrary-precision floating-point arithmetic. Different nuclear models are supported - namely the point, Gaussian, shell and uniform ball models - as well as both the RKB and DKB basis schemes. In addition every quantity is computed fully analytically, in terms of elementary or well-known special functions. It is then possible to compute vacuum polarization densities and energy shifts either with \textit{Python}, still in arbitrary precision, or by calling C routines that are implemented in double precision, with the native C type \textit{double}, quadruple precision with GCC quadmath library, or in arbitrary precision with the  MPFR library \cite{mpfr}. Each of these implementations can be selected from the \textit{Python} interface, depending on the precision needed versus computation time, along with the degree of parallelization implemented with OpenMP. Finally, as most of the computations consist of matrix products, the BLAS routines from the C interface GSL \cite{gsl} are called for efficiency. However, there is no native support for floating-point data types above double precision. To carry the matrix products in quadruple precision and beyond, we implemented the Ozaki scheme \cite{Ozaki}, which allows for efficient high floating-point precision matrix products using only a low precision underlying matrix multiplication routine.

Wichmann--Kroll energy-shifts were calculated with respect to the $\mathrm{1s}_{1/2}$-orbitals of one-electron atoms \ce{Kr^{35+}}, \ce{Xe^{53+}}, \ce{Yb^{69+}}, \ce{Pb^{81+}}, \ce{Rn^{85+}}, \ce{U^{91+}} and \ce{Mt^{108+}}. Energy-optimized basis sets for two-electron atoms
are available for the rare gases and uranium\cite{Almoukhalalati}, and we took the s-exponents from \textit{dyall\_1s2.4z}. For \ce{Yb^{69+}}, \ce{Pb^{81+}} and \ce{Mt^{108+}} we started from the list of s-exponents from the \textit{dyall.4zp} set\cite{Gomes2010,Dyall2006,Dyall2011}. However, since those basis sets are optimized for neutral atoms, we removed diffuse (small) exponents that do not contribute to more than $10^{-5}E_h$ to the ground-state energy of the one-electron atom. With the exponents sorted by increasing value, we conserved the first 18 of them for \ce{Pb^{81+}}, the first 19 exponents for \ce{Yb^{69+}}, and the exponents $0$ through $15$ and then $17$ and $18$ for \ce{Mt^{108+}}. The basis set files \textit{dyall\_1s2.4z} and \textit{dyall.4zp} are available from the DIRAC software for relativistic molecular calculations\cite{DIRAC25,DIRACpaper2020}. 

For the complete set of orbitals appearing in the VP density, we found that it was possible to generate a universal set of even-tempered exponents, that we denote \textit{wkopt}-$N$, where $N$ represents the basis size and
\beq \label{eq:wkopt}
	\zeta_{\min} = 201\,a_0^{-2} \qcomma \zeta_{\max} = 195883180777\,a_0^{-2}.
\eeq
The determination of the exponent range is explained in the next section.
 
\section{Results}\label{sec:results} 

In this section, we show the results of our energy shift calculations on different high-$Z$ hydrogen-like ions. Reference values were obtained by Soff and Mohr \cite{Soff_PRA1988}, and later Persson \textit{et al.} \cite{Persson1993} in the Green's function approach. Both compute the first partial-wave contributions, Eq.~\eqref{eq:energy_shift_partial_wave_exp}, for $\abs{\kappa}\leq 5$, a good approximation given the speed of convergence of the series. In addition, the latter performed an extrapolation of the series in $\abs{\kappa}^{-4}$ to give an estimate of the total energy shift. \\

Ivanov and co-workers repeated these calculations in even-tempered Gaussian bases, using 70-digit numbers, corresponding to 233-bit floating-point format(f233)\cite{Ivanov2024}. They provide also the individual $\abs{\kappa}$-contributions for \ce{U^{91+}} with a shell nuclear model, in both a finite Gaussian basis set and Green's function approach. These results were invaluable in carrying out the present work. 
Their results match very well those of Persson \textit{et al.}, with a relative error of less than one percent in the case of $1s_{1/2}$ orbitals. However, this was not sufficient to carry out the $\kappa$-extrapolation meaningfully, according to the authors. Here, we show how to attain the desired accuracy, and report the results of our extrapolation, comparing with Refs.~\cite{Persson1993} and \cite{Ivanov2024}. \\

\begin{figure}[!h]
\centering
\includegraphics[width=\linewidth]{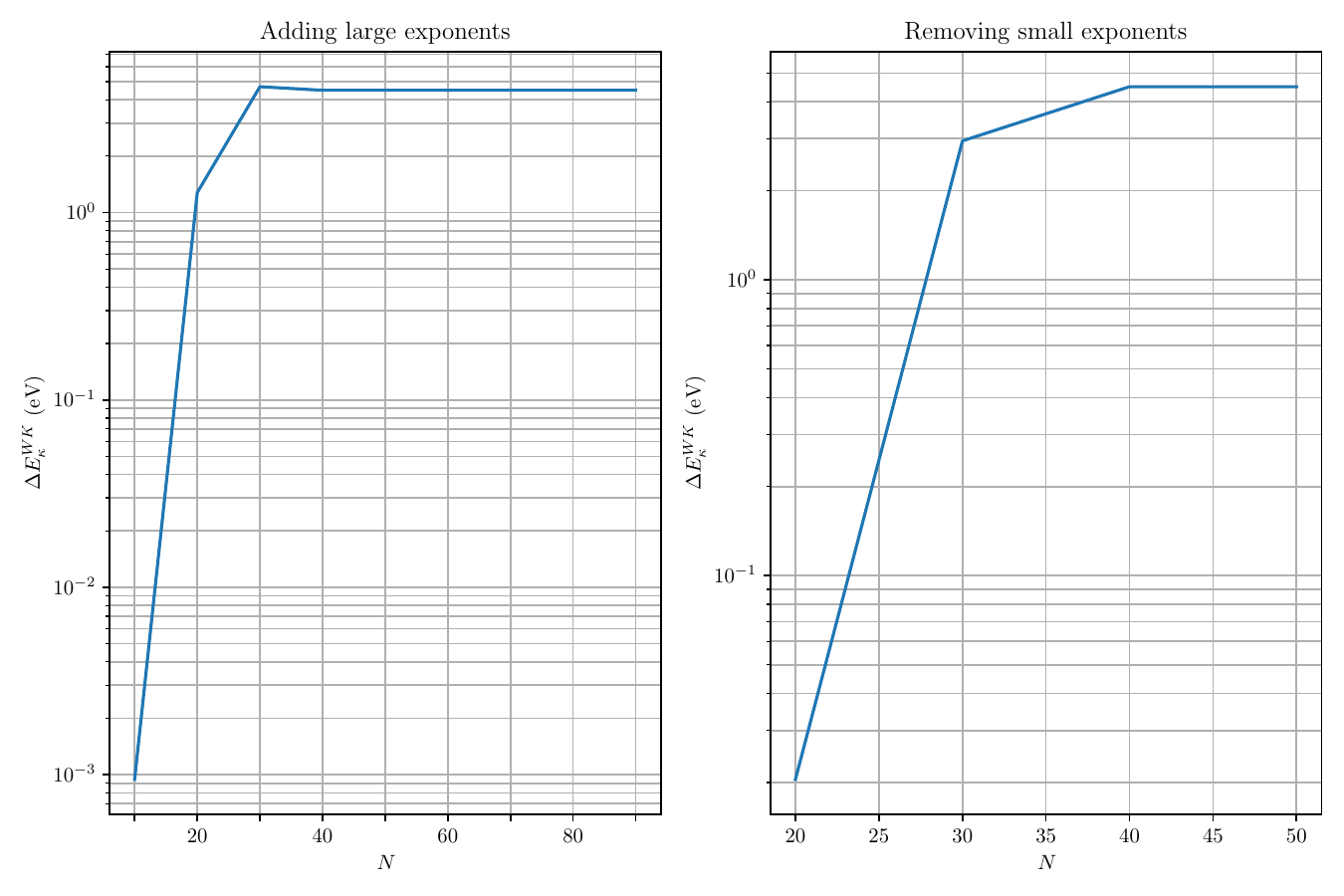}
\caption{Optimization of the vacuum even-tempered basis, performed on the $1s_{1/2}$ state of \ce{U^{91+}} for $\abs{\kappa}=1$ with a shell nucleus, $r_n=5.751\,\text{fm}$, and initial parameters $\zeta_{\min}=1\,a_0^{-2}$, $\beta=1.7$ and $N=10$. In the left plot, we added larger exponents in the even-tempered basis with $\beta$ fixed, until convergence of the energy shift. In the right plot, we then removed the smaller exponents and saw the energy shift break out of convergence. This method is what determines the suitable values of $\zeta_{\min}$ and $\zeta_{\max}$ chosen to define the basis sets \textit{wkopt}-$N$ in Eq.~\eqref{eq:wkopt}.}
\label{fig:vac_basis_optim}
\end{figure}

To describe the vacuum polarization density, a geometric sequence of exponents, Eq.~\eqref{eq:even_temp_exp}, was selected in the following manner: An even-tempered basis set is defined by the three parameters $\zeta_{\min}$, $\zeta_{\max}$, and the basis size $N$, or equivalently the parameter $\beta$. The interval of exponents $\left[\zeta_{\min}, \zeta_{\max}\right]$ determines the physical scale described by the basis set, whereas $\beta$ indicates the density of the basis. To find suitable parameters for energy shift calculations, we start by fixing $\beta$ to a reasonably dense value, but which does not create linear dependencies issues in double precision, here $\beta=1.7$ (cf. Fig.~\ref{fig:beta_select}), and selecting a small enough $\zeta_{\min}$ to be sure to capture the long-range behavior of $\rho_\kappa^{\text{WK}}$, typically $\zeta_{\min}=1\,a_0^{-2}$. We then add more and more exponents to the basis until the energy shift converges, and finally, for compactness, trim off smaller exponents until we break the convergence. Relative variation to the previous value of the energy shift is used as a convergence criterion, with a threshold of $10^{-5}$. This procedure determines the interval of exponents $\left[\zeta_{\min}, \zeta_{\max}\right]$ that is the best suited for the scale of the Wichmann--Kroll density, and we stress that it is independent of the availability of reference values or not.  The process is depicted in Fig.~\ref{fig:vac_basis_optim} in the case of the shell nucleus \ce{U^{91+}} for $\abs{\kappa}=1$. 

\begin{figure}[!h]
\centering
\includegraphics[width=\linewidth]{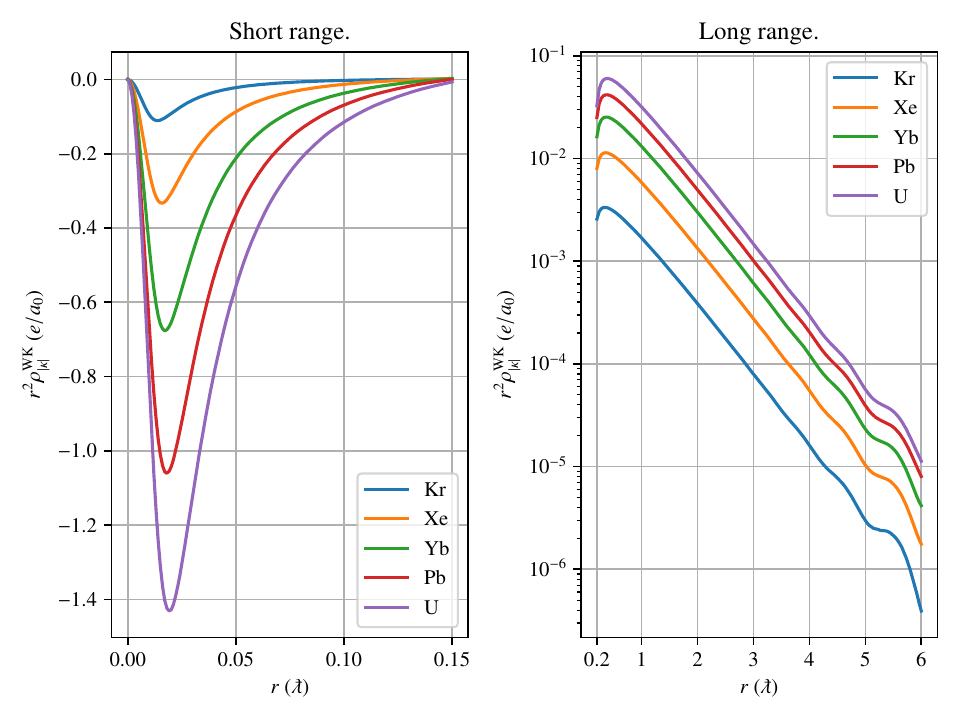}
\caption{Wichmann-Kroll vacuum polarization densities for $Z = 36, 54, 70, 82, 92$ and $\abs{\kappa}=1$. Computed with a uniform nucleus with $r_n=4.230,4.826,5.273,5.505,5.860\,\text{fm}$ respectively and the basis \textit{wkopt}-60 ($\beta = 1.42$), see Eq.~\eqref{eq:wkopt}.}
\label{fig:wk_densities}
\end{figure}

\begin{figure}[!h]
\centering
\includegraphics[width=\linewidth]{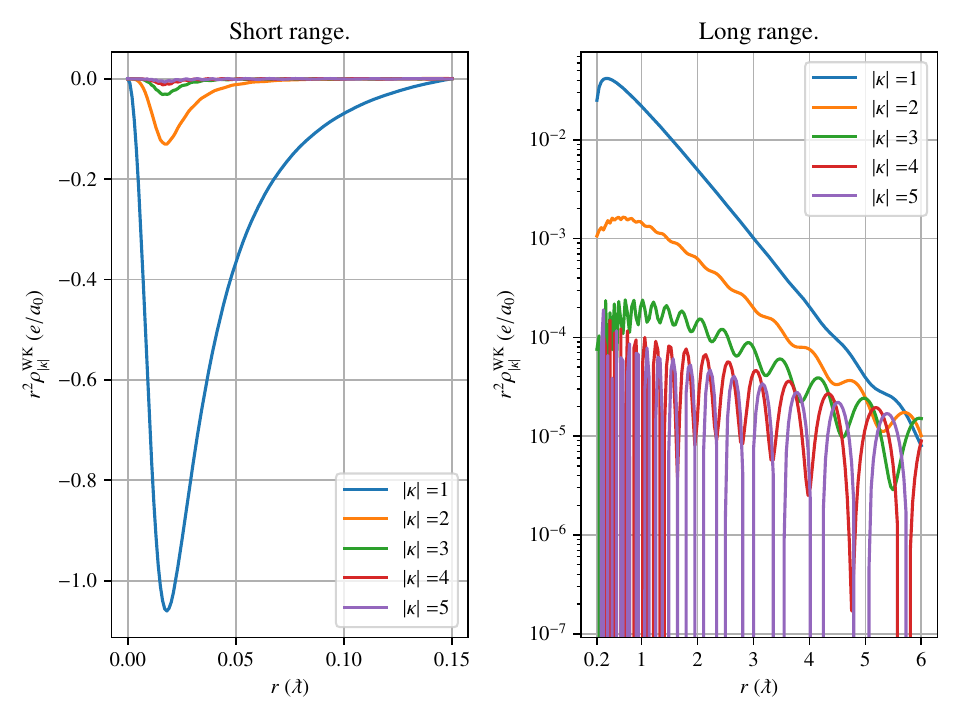}
\caption{Wichmann--Kroll vacuum polarization densities for $Z = 82$ and $\abs{\kappa}=1,\dots,5$. Computed with a uniform nucleus with $r_n=5.505\,\text{fm}$ and the basis \textit{wkopt}-60 ($\beta = 1.42$), see Eq.~\eqref{eq:wkopt}. The oscillations at high $\kappa$ seems to be a defect of the finite basis representation, they can be only partially damped for larger basis size (typically $N=120$).}
\label{fig:wk_kappa_densities}
\end{figure}

One would \emph{a priori} expect having to generate even-tempered basis sets for each nuclear charge $Z$ and for each $\abs{\kappa}$, which would be quite a bit of work. However, from Fig.~\ref{fig:wk_densities} we see that the WK densities ($\abs{\kappa}=1$) for the atoms under consideration have the same spatial localization. It was indeed already known to Wichmann and Kroll that the mean radius of $\rho^{\text{WK}}$ is only weakly dependent on $Z\alpha$ \cite{WichmannKroll}. A similar observation can be made from Fig.~\ref{fig:wk_kappa_densities} for the different $\abs{\kappa}$-contributions. This means that the same basis can be used for all of our energy shift calculations. The resulting universal set \textit{wkopt}-$N$ is specified in Section \ref{sec:comp}.

\begin{figure}[!h]
\centering
\includegraphics[width=\linewidth]{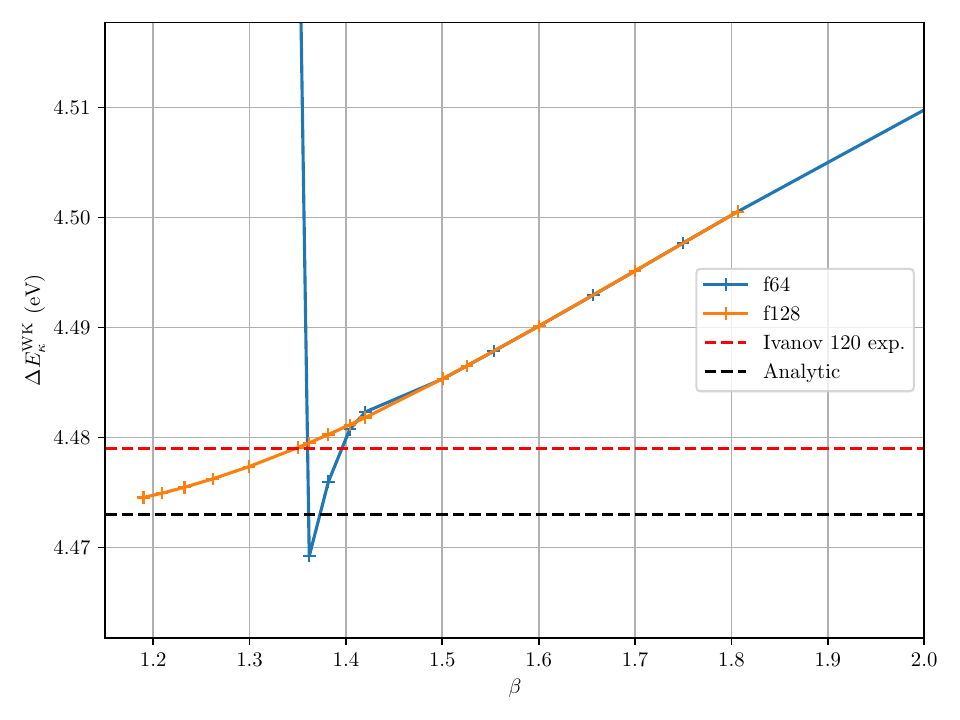}
\caption{Optimization of the vacuum even-tempered basis \textit{wkopt}-$N$ by varying $N$, hence $\beta$, for the calculation of the Wichmann--Kroll energy shift on the shell nucleus \ce{U^{91+}} with $r_n=5.751\,\text{fm}$. The black and red dotted lines denotes respectively the Green's function integration and the best Gaussian basis ($N=120, \beta=1.17$) results of Ivanov \textit{et al.} \cite{Ivanov2024}.}
\label{fig:wk_eshift_conv}
\end{figure}

In Table \ref{tab:U_shell_eshift}, we report our calculated WK-energy shifts of the $1s_{1/2}$ orbital of \ce{U^{91+}}, with a shell nuclear model with $r_n=5.751\,\text{fm}$. We also report the reference values obtained by Ivanov \textit{et al.}  in Gaussian basis and with their Green's function integration method\cite{Ivanov2024}. It can be seen that our results in double-precision using \textit{wkopt}-60, corresponding to $\beta=1.42$ can not compete with the results obtained by Ivanov \textit{et al.} with their best Gaussian basis ($N=120, \beta=1.17$) and, as already mentioned, using 70-digit numbers (f233)\cite{Ivanov2024}. More precise results can be attained with a denser basis by reducing $\beta$. However, as seen in Fig.~\ref{fig:wk_eshift_conv}, numerical noise becomes an issue in double precision calculations around $\beta=1.4$. We therefore switch to quadruple precision, allowing us to use \textit{wkopt}-120, corresponding to $\beta=1.19$.  We see that with this basis the error is reduced to $1.6\times10^{-3}E_h$ for $\abs{\kappa}=1$, but the higher $\kappa$-contributions are more difficult to converge.

The obtained precision is already quite satisfying, but there is still room for improvement. Increasing the floating-point precision significanly increases the computational cost. We have therefore resorted to extrapolation. A nice feture using even-tempered basis sets is that the complete basis set limit can be straightforwardly attained using Eq.~\eqref{eq:ebas_lim}. We have already established an interval $\left[\zeta_{\min}, \zeta_{\max}\right]$ by a procedure for which we can expect that the contribution from exponents outside of this range can be neglected. The complete basis set limit can thereby be very precisely approximated by solely taking the limit $\beta \rightarrow 1^{+}$.

\begin{figure}[!h]
\centering
\includegraphics[width=\linewidth]{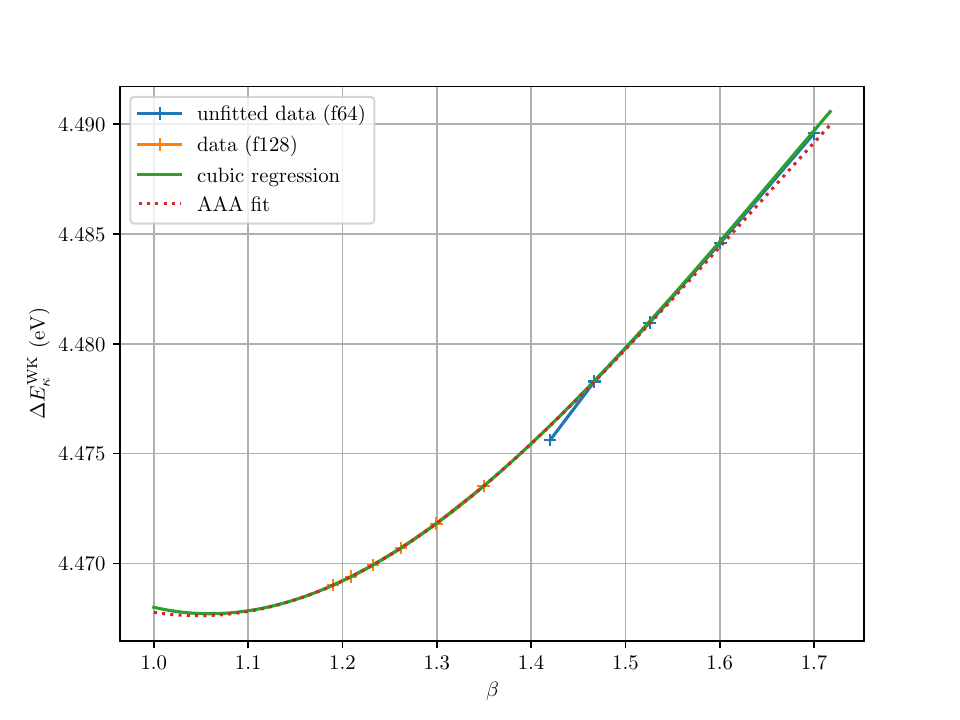}
\caption{Cubic polynomial and AAA fit of the $\abs{\kappa}=1$ contribution to the Wichmann--Kroll energy shift for the uniform ball nucleus \ce{U^{91+}} with $r_n=5.860\,\text{fm}$ and the basis \textit{wkopt}-$N$. The fitted data corresponds to quadruple (f128) precision calculations, and the unfitted one to double (f64) precision.}
\label{fig:wk_eshift_comp_basis_fit}
\end{figure}

\begin{figure}[!h]
\centering
\includegraphics[width=\linewidth]{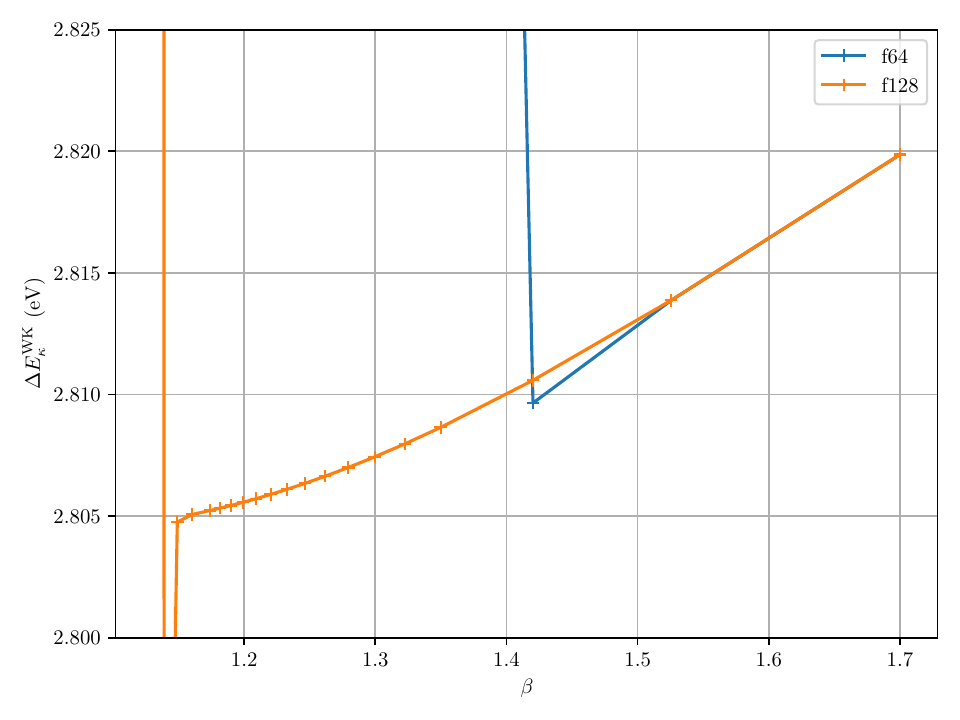}
\caption{Convergence of the WK energy shift in \ce{Rn^{85+}} for $\abs{\kappa}=1$ using the \textit{wkopt}-$N$ basis sets and a uniformly charged ball nuclear model with $r_n=5.632\,\text{fm}$ \cite{Visscher1997}. Computations are performed in double (f64) and quadruple (f128) precisions. We observe a sudden degradation of the double precision calculations around $\beta=1.4$, and a similar phenomenon for quadruple precision around $\beta=1.1$, which is the sign of numerical noise as described in sec.\ref{sec:numerical_noise}.}
\label{fig:wk_eshift_Rn}
\end{figure}

Each $\kappa$-contribution to the Wichmann--Kroll energy shift can be estimated in the complete basis set limit by an extrapolation of the computed data. This procedure, however, induces an uncertainty on the extrapolated value, which we would like to be as small as possible. To quantify this uncertainty, we have found that comparing two different extrapolation methods works best. We perform a cubic polynomial regression and an AAA interpolation \cite{AAA} on the available data while trying to avoid overfitting. 
\beq
	\begin{split}
		f^{\text{cubic}}(\beta) &= a\beta^3 + b\beta^2 + c\beta + d \\
		f^{\text{AAA}}(\beta) &= \left.\sum_{j=1}^{m} \frac{w_j f_j}{\beta - \beta_j}\right/ \sum_{j=1}^m \frac{w_j}{\beta - \beta_j}.
	\end{split}
\eeq
In the case of the AAA-fit, we denote by $m$ the number of data points, $\beta_j$ the computed beta values, $f_j$ the corresponding values for $\Delta E_\abs{\kappa}^{\text{WK}}(\beta_j)$ and $w_j$ the AAA-weights. We note that the cubic regression is in principle only valid for our extrapolation in a small neighborhood around $\beta=1$, whereas the AAA approximants allow for a better description at larger $\beta$. We then want to select enough points so that the fits interpolates our data precisely, with a low covariance of the fitting parameters, but not too much or too large ones so that the fit uncertainty, that we define as $\abs{f^{\text{cubic}}(1) - f^{\text{AAA}}(1)}$, remains below the desired precision of $10^{-4}E_h$. We see for instance in Fig.~\ref{fig:wk_eshift_comp_basis_fit} that with suitable data, we can considerably reduce this fit uncertainty, while also maintaining sufficient precision on the unfitted data. Interestingly, we note that the fits are never strictly monotonous, with a minimum often reached below $\beta=1.1$. This poses a certain difficulty for finding a suitable fit as we were not able to reach $\beta=1.1$ because of numerical noise, even in quadruple precision, which is supported by the analysis following Eq.~\eqref{eq:NR_VP_bound} and the calculations reported in Fig.~\ref{fig:wk_eshift_Rn} on \ce{Rn^{85+}}. The small fit uncertainty, however, gives us confidence in our extrapolation method. One observes in Table \ref{tab:U_shell_eshift} that the extrapolation provides perfect agreement with the reference data obtained with the Green's function approach. We have tried to obtain similar precision extrtapolating from double-precision data, but so far without success.

\begin{table*}
  \caption{Calculations of $\Delta E_\kappa^{\text{WK}}$ (in eV) of the $1s_{1/2}$ orbital of \ce{U^{91+}}, using a shell nuclear model with $r_n=5.751\,\text{fm}$. We report double- (f64) and quadruple-precision (f128) calculations in \textit{wkopt-}$N$ bases, cubic polynomial regression and AAA-approximation extrapolations to the complete basis set limit $\beta=1$, along with the analytic computation reference results of Ivanov \textit{et al.} \cite{Ivanov2024} and their best result using 120 Gaussian basis functions. We also indicate the fit uncertainty, defined as $\abs{f^{\text{cubic}}(1) - f^{\text{AAA}}(1)}$ and reported in eV.}
  \label{tab:U_shell_eshift}
  \begin{ruledtabular}
  \begin{tabular}{cccccccc}
    $\abs{\kappa}$ & Green int. \cite{Ivanov2024} &  f233($N=120$)\cite{Ivanov2024} & f64 ($N=60$) & f128 ($N=120$) & Cubic extr. & AAA extr. & Fit uncertainty \\
    \hline
    1 & 4.473 & 4.479 & 4.4823 & 4.4746 & 4.4733 & 4.4734 & $1.0\times10^{-4}$ \\
    2 & 0.394 & 0.396 & 0.4082 & 0.3964 & 0.3931 & 0.3939 & $7.6\times10^{-4}$ \\
    3 & 0.081 & 0.085 & 0.0993 & 0.0852 & 0.0810 & 0.0806 & $4.6\times10^{-4}$ \\
    4 & 0.024 & 0.029 & 0.0441 & 0.0294 & 0.0242 & 0.0246 & $4.3\times10^{-4}$ \\
    5 & 0.009 & 0.014 & 0.0289 & 0.0147 & 0.0094 & 0.0092 & $2.5\times10^{-4}$ \\
  \end{tabular}
  \end{ruledtabular}
    \vspace{-4mm}
\end{table*}

After performing this extrapolation to the complete basis limit for every $Z$ and $\abs{\kappa}$-contribution, and following Persson \textit{et al.}, we can finally perform the extrapolations in $\abs{\kappa}^{-4}$ of the partial-wave expansions. We consider the mean of the cubic and AAA extrapolations as our result for each $\kappa$-contribution, and accumulate the fit uncertainties to give an estimate of the total one. The results are displayed in table \ref{tab:eshift_results}. 

\begin{table*}
  \caption{Calculations of $\Delta E_{\kappa}^{\text{WK}}$ for $\abs{\kappa}\leq5$ and extrapolation of the partial wave series for different atoms. These calculations are performed with a uniformly charged ball nucleus. Dyall $4$z bases are used for the reference orbitals, and for the vacuum even-tempered the \textit{wkopt}-$N$ bases with $N$ chosen such that the fit uncertainty $\abs{f^{\text{cubic}}(1) - f^{\text{AAA}}(1)}$ for each $\kappa$-contribution is of the order of $10^{-4}\,\text{eV}$ at worse. 
    Note that the total fit uncertainty reported here is the sum of these uncertainties for each $\kappa$-contribution, the uncertainty on the partial wave series extrapolation is not evaluated.
    Energies and uncertainties are reported in eV. }
  \label{tab:eshift_results}
  \begin{ruledtabular}
  \begin{tabular}{lccccc}
    Element & $r_n$ (fm) \cite{Persson1993} & Complete basis extr. ($\abs{\kappa}\leq5$) & Total fit uncert. & $\kappa$-extrapolation & Ref. \cite{Persson1993} \\
    \hline
    Kr & $4.230$ & 0.015428 & $1.4\times10^{-5}$ & 0.015479 & 0.0155 \\
    Xe & $4.826$ & 0.168893 & $1.0\times10^{-4}$ & 0.169410 & 0.1695 \\
    Yb & $5.273$ & 0.826520 & $2.0\times10^{-4}$ & 0.828855 & 0.8283 \\
    Pb & $5.505$ & 2.281762 & $2.7\times10^{-4}$ & 2.287805 & 2.2900 \\
    Rn & $5.632$ & 3.139294 & $5.5\times10^{-4}$ & 3.145348 & -- \\
    U & $5.860$ & 4.978143 & $5.0\times10^{-4}$ & 4.985659 & 4.9863 \\
    Mt & $5.947$ & 17.583126 & $4.5\times10^{-3}$ & 17.604984 & --
  \end{tabular}
  \end{ruledtabular}
    \vspace{-4mm}
\end{table*} 
`
\section{Conclusion and perspectives}\label{sec:conc}
 
In the present work we have established a framework for the computation of one-loop vacuum polarization effects in one-electron ions using finite
Gaussian basis sets. We have discussed and proposed various formulations of the vacuum polarization density and their relation. 
The energy shift associated with vacuum polarization
is formulated as the contraction of two-electron integrals with two density matrices, one representing the reference orbital, the other the VP density. 
With the use of Riesz projectors an analytic expression for the effective density matrix representing the divergent linear contribution to the VP density
has been formulated, valid in any finite basis approximation. The non-linear
Wichmann--Kroll contribution is then obtained by subtracting the linear part from
the total VP density matrix.

The reference orbital has been expanded in an energy-optimized basis, whereas
the VP density is generated from the complete set of solutions to the Dirac
problem in an even-tempered basis. We observe that the non-linear VP density
has basically the same spatial localization for all nuclear charges $Z$ and values of $\kappa$, allowing the formulation of a universal even-tempered VP basis set that we denote \textit{wkopt}-$N$. It is characterized by a fixed exponent
interval $\left[\zeta_{\min}, \zeta_{\max}\right]$ and can be make increasingly dense by increasing the number $N$ of functions in this interval.

In the course of calculations however, linear dependencies became an issue, so we have carried out an extensive analysis of the numerical sensitivity of the model. We establish a relation between linear dependence and the choice of the density parameter $\beta$. We report a quadruple-precision implementation that allows us to push harder towards the complete-basis limit. For this purpose the Ozaki scheme was implemented to perform matrix products efficiently in quadruple precision, using the lower precision \textit{DGEMM} routine. In practice we find that numerical differences occur at $\beta=1.4$ and $\beta=1.19$ in double and quadruple precision, respectively, and provide theoretical justification for these observations.

We find that it is possible to go beyond quadruple precision and obtain very accurate results for individual $\kappa$-contributions, comparable to the Green's function approach of \cite{Soff_PRA1988, Persson1993, Ivanov2024},  by extrapolation towards $\beta = 1$. It was furthermore possible to report total energy shift estimations by also extrapolating the partial-wave series in $\abs{\kappa}^{-4}$. The results are in very good agreement with \cite{Persson1993}, with errors and uncertainties on the order of $10^{-3}\,\text{eV}$. We believe that is possible to improve precision by further refinement of the extrapolation procedure.

Together with Refs. \cite{Ivanov2024, Ferenc2025} the present work demonstrte that Gaussian finite basis sets are able to give a full account of one-loop QED effects in hydrogen-like ions. It would certainly be interesting to encompass also the linear contribution to the VP density, the Uehling term, in the same framework. A regularization and renormalization approach for use with finite Gaussian bases is in progress. 
The next natural step should be the consideration of more complex systems, like multi-electron atoms or two-center molecular systems. We expect the Gaussian basis approach to perform very well in this situation too.

\begin{acknowledgments}
The authors thank D{\'a}vid Ferenc, Jan Brandejs, Anthony Scemama and Vladislav Ivanov for helpful discussions. This project was funded by the European Research Council (ERC) under the European Union’s Horizon 2020 research and innovation programme (Grant Agreement No. ID:101019907).
\end{acknowledgments}

\appendix


\newpage

\bibliography{biblio.bib}

@article{WichmannKroll,
  title = {{Vacuum Polarization in a Strong Coulomb Field}},
  author = {Wichmann, Eyvind H. and Kroll, Norman M.},
  journal = {Phys. Rev.},
  volume = {101},
  issue = {2},
  pages = {843--859},
  numpages = {0},
  year = {1956},
  month = {Jan},
  publisher = {American Physical Society},
  doi = {10.1103/PhysRev.101.843},
}

@article{PauliRose,
  title = {{Remarks on the Polarization Effects in the Positron Theory}},
  author = {Pauli, W. and Rose, M. E.},
  journal = {Phys. Rev.},
  volume = {49},
  issue = {6},
  pages = {462--465},
  numpages = {0},
  year = {1936},
  month = {Mar},
  publisher = {American Physical Society},
  doi = {10.1103/PhysRev.49.462},
}

@article{Uehling,
  title = {{Polarization Effects in the Positron Theory}},
  author = {Uehling, E. A.},
  journal = {Phys. Rev.},
  volume = {48},
  issue = {1},
  pages = {55--63},
  numpages = {0},
  year = {1935},
  month = {Jul},
  publisher = {American Physical Society},
  doi = {10.1103/PhysRev.48.55},
}

@article{Peierls1934,
  title={{The vacuum in Dirac's theory of the positive electron}},
  author={Peierls, R.},
  journal={Proc. R. Soc. Lond. Series A},
  volume={146},
  number={857},
  pages={420--441},
  year={1934},
  publisher={The Royal Society London},
  doi={10.1098/rspa.1934.0164}
}

@article{Serber1935,
  title = {{Linear Modifications in the Maxwell Field Equations}},
  author = {Serber, R.},
  journal = {Phys. Rev.},
  volume = {48},
  issue = {1},
  pages = {49--54},
  numpages = {0},
  year = {1935},
  month = {Jul},
  publisher = {American Physical Society},
  doi = {10.1103/PhysRev.48.49},
}

@article{Schwinger1949II,
  title = {{Quantum Electrodynamics. II. Vacuum Polarization and Self-Energy}},
  author = {Schwinger, Julian},
  journal = {Phys. Rev.},
  volume = {75},
  issue = {4},
  pages = {651--679},
  numpages = {0},
  year = {1949},
  month = {Feb},
  publisher = {American Physical Society},
  doi = {10.1103/PhysRev.75.651}
}

@article{KarplusNeuman,
  title = {{Non-Linear Interactions between Electromagnetic Fields}},
  author = {Karplus, Robert and Neuman, Maurice},
  journal = {Phys. Rev.},
  volume = {80},
  issue = {3},
  pages = {380--385},
  numpages = {0},
  year = {1950},
  month = {Nov},
  publisher = {American Physical Society},
  doi = {10.1103/PhysRev.80.380}
}

@article{Schwarz1982,
author = {W.H.E. Schwarz and H. Wallmeier},
title = {{Basis set expansions of relativistic molecular wave equations}},
journal = {Mol. Phys.},
volume = {46},
number = {5},
pages = {1045--1061},
year = {1982},
publisher = {Taylor \& Francis},
doi = {10.1080/00268978200101771}
}

@article{Ishikawa1987,
author = {Ishikawa, Yasuyuki and Quiney, H. M.},
title = {{On the use of an extended nucleus in Dirac–Fock Gaussian basis set calculations}},
journal = {Int. J. Quantum Chem.},
volume = {32},
number = {S21},
pages = {523-532},
doi = {10.1002/qua.560320751},
year = {1987}
}

@article{Shabaev2004,
  title={{Dual kinetic balance approach to basis-set expansions for the Dirac equation}},
  author={Shabaev, Vladimir M and Tupitsyn, Ilya I and Yerokhin, Vladimir A and Plunien, G{\"u}nter and Soff, G},
  journal={Phys. Rev. Lett.},
  volume={93},
  number={13},
  pages={130405},
  year={2004},
  doi={10.1103/PhysRevLett.93.130405},
  publisher={APS}
}

@article{Johnson1985,
title = {{The Lamb shift in hydrogen-like atoms, 1 $\leq Z \leq$ 110}},
journal = {At. Data Nucl. Data Tables},
volume = {33},
number = {3},
pages = {405-446},
year = {1985},
issn = {0092-640X},
doi = {10.1016/0092-640X(85)90010-5},
author = {W.R. Johnson and Gerhard Soff}
}

@Article{Yerokhin2020,
AUTHOR = {Yerokhin, Vladimir A. and Maiorova, Anna V.},
TITLE = {{Calculations of QED Effects with the Dirac Green Function}},
JOURNAL = {Symmetry},
VOLUME = {12},
YEAR = {2020},
NUMBER = {5},
ARTICLE-NUMBER = {800},
ISSN = {2073-8994},
doi = {10.3390/sym12050800}
}

@book{Grant2007,
  editor    = {Grant, I. P.},
  title     = {{Relativistic Quantum Theory of Atoms and Molecules: Theory and Computation}},
  publisher = {Springer},
  address   = {New York, NY},
  year      = {2007},
  series    = {Springer Series on Atomic, Optical, and Plasma Physics},
  edition   = {1},
  isbn      = {978-0-387-35069-1},
  doi       = {10.1007/978-0-387-35069-1}
}

@article{Blomqvist1972,
title = {{Vacuum polarization in exotic atoms}},
journal = {Nucl. Phys. B},
volume = {48},
number = {1},
pages = {95-103},
year = {1972},
issn = {0550-3213},
doi = {10.1016/0550-3213(72)90051-X},
author = {J. Blomqvist}
}

@article{RinkerWilets1973,
  title = {{Vacuum Polarization in High-$Z$, Finite-Size Nuclei}},
  author = {Rinker, G. A. and Wilets, L.},
  journal = {Phys. Rev. Lett.},
  volume = {31},
  issue = {26},
  pages = {1559--1562},
  numpages = {0},
  year = {1973},
  month = {Dec},
  publisher = {American Physical Society},
  doi = {10.1103/PhysRevLett.31.1559}
}

@article{RinkerWilets1975,
  title = {{Vacuum polarization in strong, realistic electric fields}},
  author = {Rinker, G. A. and Wilets, L.},
  journal = {Phys. Rev. A},
  volume = {12},
  issue = {3},
  pages = {748--762},
  numpages = {0},
  year = {1975},
  month = {Sep},
  publisher = {American Physical Society},
  doi = {10.1103/PhysRevA.12.748}
}

@article{Gyulassy1974Muonic,
  title = {{Nuclear-Size Effects on Vacuum Polarization in Muonic Pb}},
  author = {Gyulassy, M.},
  journal = {Phys. Rev. Lett.},
  volume = {32},
  issue = {24},
  pages = {1393--1396},
  numpages = {0},
  year = {1974},
  month = {Jun},
  publisher = {American Physical Society},
  doi = {10.1103/PhysRevLett.32.1393}
}

@article{Gyulassy1974Collisions,
  title = {{Vacuum Polarization in Heavy-Ion Collisions}},
  author = {Gyulassy, Miklos},
  journal = {Phys. Rev. Lett.},
  volume = {33},
  issue = {15},
  pages = {921--925},
  numpages = {0},
  year = {1974},
  month = {Oct},
  publisher = {American Physical Society},
  doi = {10.1103/PhysRevLett.33.921}
}

@article{Gyulassy1975,
title = {{Higher order vacuum polarization for finite radius nuclei}},
journal = {Nucl. Phys. A},
volume = {244},
number = {3},
pages = {497-525},
year = {1975},
issn = {0375-9474},
doi = {10.1016/0375-9474(75)90554-0},
author = {Miklos Gyulassy}
}

@article{Neghabian1983,
  title = {{Vacuum polarization for an electron in a strong Coulomb field}},
  author = {Neghabian, A. R.},
  journal = {Phys. Rev. A},
  volume = {27},
  issue = {5},
  pages = {2311--2320},
  numpages = {0},
  year = {1983},
  month = {May},
  publisher = {American Physical Society},
  doi = {10.1103/PhysRevA.27.2311}
}

@article{Mohr1998,
  title={{QED corrections in heavy atoms}},
  author={Mohr, Peter J and Plunien, G{\"u}nter and Soff, Gerhard},
  journal={Phys. Rep.},
  volume={293},
  number={5-6},
  pages={227--369},
  year={1998},
  publisher={Elsevier},
  doi={10.1016/S0370-1573(97)00046-X}
}

@article{Soff_PRA1988,
  title = {{Vacuum polarization in a strong external field}},
  author = {Soff, Gerhard and Mohr, Peter J.},
  journal = {Phys. Rev. A},
  volume = {38},
  issue = {10},
  pages = {5066--5075},
  numpages = {0},
  year = {1988},
  month = {Nov},
  publisher = {American Physical Society},
  doi = {10.1103/PhysRevA.38.5066},
}

@article{Persson1993,
  title = {{Accurate vacuum-polarization calculations}},
  author = {Persson, Hans and Lindgren, Ingvar and Salomonson, Sten and Sunnergren, Per},
  journal = {Phys. Rev. A},
  volume = {48},
  issue = {4},
  pages = {2772--2778},
  numpages = {0},
  year = {1993},
  month = {Oct},
  publisher = {American Physical Society},
  doi = {10.1103/PhysRevA.48.2772},
}

@article{Ivanov2024,
  title = {{Vacuum-polarization Wichmann-Kroll correction in the finite-basis-set approach}},
  author = {Ivanov, V. K. and Baturin, S. S. and Glazov, D. A. and Volotka, A. V.},
  journal = {Phys. Rev. A},
  volume = {110},
  issue = {3},
  pages = {032815},
  numpages = {11},
  year = {2024},
  month = {Sep},
  publisher = {American Physical Society},
  doi = {10.1103/PhysRevA.110.032815}
}

@article{Indelicato2014,
  title = {{Coordinate-space approach to vacuum polarization}},
  author = {Indelicato, Paul and Mohr, Peter J. and Sapirstein, J.},
  journal = {Phys. Rev. A},
  volume = {89},
  issue = {4},
  pages = {042121},
  numpages = {10},
  year = {2014},
  month = {Apr},
  publisher = {American Physical Society},
  doi = {10.1103/PhysRevA.89.042121}
}

@article{Sommerfeldt2025,
  title={{All-Order Wichmann and Kroll Contribution in Heavy Electronic and Exotic Atoms}},
  author={Sommerfeldt, Jonas and Indelicato, Paul},
  journal={arXiv preprint arXiv:2509.08763},
  year={2025},
  eprint={2509.08763},
  archivePrefix={arXiv},
  primaryClass={physics.atom-ph}
}

@article{Salman_PRA2023,
  title = {{Calculating the many-potential vacuum polarization density of the Dirac equation in the finite-basis approximation}},
  author = {Salman, Maen and Saue, Trond},
  journal = {Phys. Rev. A},
  volume = {108},
  issue = {1},
  pages = {012808},
  numpages = {13},
  year = {2023},
  month = {Jul},
  publisher = {American Physical Society},
  doi = {10.1103/PhysRevA.108.012808}
}

@article{Ferenc2025,
  title = {{Gaussian-basis-set approach to one-loop self-energy}},
  author = {Ferenc, D\'avid and Salman, Maen and Saue, Trond},
  journal = {Phys. Rev. A},
  volume = {111},
  issue = {4},
  pages = {L040802},
  numpages = {6},
  year = {2025},
  month = {Apr},
  publisher = {American Physical Society},
  doi = {10.1103/PhysRevA.111.L040802}
}

@phdthesis{SalmanThesis,
  TITLE = {{Quantum Electrodynamic Corrections in Quantum Chemistry}},
  AUTHOR = {Salman, Maen},
  URL = {https://theses.hal.science/tel-03715663},
  NUMBER = {2022TOU30032},
  SCHOOL = {{Universit{\'e} Paul Sabatier - Toulouse III}},
  YEAR = {2022},
  MONTH = Jan,
  TYPE = {Theses},
  HAL_ID = {tel-03715663},
  HAL_VERSION = {v1},
}

@article{Chaix1989I,
doi = {10.1088/0953-4075/22/23/004},
year = {1989},
month = {dec},
publisher = {},
volume = {22},
number = {23},
pages = {3791},
author = {P Chaix and D Iracane},
title = {{From quantum electrodynamics to mean-field theory. I. The Bogoliubov-Dirac-Fock formalism}},
journal = {J. Phys. B}
}

@article{Chaix1989II,
doi = {10.1088/0953-4075/22/23/005},
year = {1989},
month = {dec},
publisher = {},
volume = {22},
number = {23},
pages = {3815},
author = {P Chaix and D Iracane and P L Lions},
title = {{From quantum electrodynamics to mean-field theory. II. Variational stability of the vacuum of quantum electrodynamics in the mean-field approximation}},
journal = {J. Phys. B}
}

@article{HainzlSiedentop,
  title={{Non-perturbative mass and charge renormalization in relativistic no-photon quantum electrodynamics}},
  author={Hainzl, Christian and Siedentop, Heinz},
  journal={Commun. Math. Phys.},
  volume={243},
  number={2},
  pages={241--260},
  year={2003},
  doi={10.1007/s00220-003-0958-6},
  publisher={Springer}
}

@article{Barbaroux2005,
  author  = {Barbaroux, Jean-Marie and Farkas, Walter and Helffer, Bernard and Siedentop, Heinz},
  title   = {{On the Hartree-Fock Equations of the Electron-Positron Field}},
  journal = {Commun. Math. Phys.},
  volume  = {255},
  number  = {1},
  pages   = {131--159},
  year    = {2005},
  month   = {apr},
  doi     = {10.1007/s00220-004-1156-x},
  issn    = {1432-0916}
}

@article{Hainzl2005Existence,
  author  = {Hainzl, Christian and Lewin, Mathieu and S\'er\'e, \'Eric},
  title   = {{Existence of a Stable Polarized Vacuum in the Bogoliubov-Dirac-Fock Approximation}},
  journal = {Commun. Math. Phys.},
  volume  = {257},
  number  = {3},
  pages   = {515--562},
  year    = {2005},
  month   = {aug},
  doi     = {10.1007/s00220-005-1343-4}
}

@article{Hainzl2005Self,
doi = {10.1088/0305-4470/38/20/014},
year = {2005},
month = {may},
publisher = {},
volume = {38},
number = {20},
pages = {4483},
author = {Hainzl, Christian and Lewin, Mathieu and S\'er\'e, \'Eric},
title = {{Self-consistent solution for the polarized vacuum in a no-photon QED model}},
journal = {J. Phys. A}
}

@article{Hainzl2007,
  title = {{Minimization method for relativistic electrons in a mean-field approximation of quantum electrodynamics}},
  author = {Hainzl, Christian and Lewin, Mathieu and S\'er\'e, Eric and Solovej, Jan Philip},
  journal = {Phys. Rev. A},
  volume = {76},
  issue = {5},
  pages = {052104},
  numpages = {11},
  year = {2007},
  month = {Nov},
  publisher = {American Physical Society},
  doi = {10.1103/PhysRevA.76.052104}
}

@article{Gravejat2009,
  author  = {Gravejat, Philippe and Lewin, Mathieu and S\'er\'e, \'Eric},
  title   = {{Ground State and Charge Renormalization in a Nonlinear Model of Relativistic Atoms}},
  journal = {Commun. Math. Phys.},
  volume  = {286},
  number  = {1},
  pages   = {179--215},
  year    = {2009},
  month   = {feb},
  doi     = {10.1007/s00220-008-0660-9},
  issn    = {1432-0916}
}

@article{Gravejat2011,
  author  = {Gravejat, Philippe and Lewin, Mathieu and S\'er\'e, \'Eric},
  title   = {{Renormalization and Asymptotic Expansion of Dirac's Polarized Vacuum}},
  journal = {Commun. Math. Phys.},
  volume  = {306},
  number  = {1},
  pages   = {1--33},
  year    = {2011},
  month   = {aug},
  doi     = {10.1007/s00220-011-1271-4},
  issn    = {1432-0916}
}

@inbook{Lewin2010,
author = {Mathieu Lewin},
title = {{Renormalization of Dirac's Polarized Vacuum}},
booktitle = {Mathematical Results in Quantum Physics},
chapter = {},
pages = {45-59},
doi = {10.1142/9789814350365_0004},
}

@book{Lewin2022,
  title={{Th{\'e}orie spectrale et m{\'e}canique quantique}},
  author={Lewin, Mathieu},
  volume={87},
  year={2022},
  publisher={Springer Cham},
  series    = {{Mathématiques et Applications}},
  isbn      = {978-3-030-93436-1},
  doi       = {10.1007/978-3-030-93436-1}
}

@article{Kato1949,
    author = {Kato, Tosio},
    title = {{On the Convergence of the Perturbation Method. I}},
    journal = {Prog. Theor. Phys.},
    volume = {4},
    number = {4},
    pages = {514-523},
    year = {1949},
    month = {12},
    issn = {0033-068X},
    doi = {10.1143/ptp/4.4.514}
}

@book{Kato,
  title={{Perturbation Theory for Linear Operators}},
  author={Kato, Tosio},
  publisher = {Springer},
  address   = {Berlin, Heidelberg},
  year      = {2012},
  series    = {Classics in Mathematics},
  edition   = {2},
  isbn      = {978-3-642-66282-9},
  doi       = {10.1007/978-3-642-66282-9},
  note      = {Originally published as volume 132 in the series: Grundlehren der mathematischen Wissenschaften}
}

@book{ReedSimon,
  title={Methods of modern mathematical physics: Functional analysis},
  author={Reed, Michael and Simon, Barry},
  volume={1},
  year={1980},
  publisher={Gulf Professional Publishing}
}

@book{Thaller,
  title={{The Dirac equation}},
  author={Thaller, Bernd},
  year={2010},
  publisher={Springer},
  address = {Berlin, Heidelberg},
  series = {Theoretical and Mathematical Physics},
  edition = {1},
  isbn = {978-3-642-08134-7},
  doi = {10.1007/978-3-662-02753-0}
}

@book{Higham,
author = {Higham, Nicholas J.},
title = {{Accuracy and Stability of Numerical Algorithms}},
publisher = {{Society for Industrial and Applied Mathematics}},
year = {2002},
doi = {10.1137/1.9780898718027},
address = {},
edition   = {Second}
}

@article{Blanchard,
  title={{Mixed precision block fused multiply-add: Error analysis and application to GPU tensor cores}},
  author={Blanchard, Pierre and Higham, Nicholas J and Lopez, Florent and Mary, Theo and Pranesh, Srikara},
  journal={SIAM J. Sci. Comput.},
  volume={42},
  number={3},
  pages={C124--C141},
  year={2020},
  doi = {10.1137/19M1289546}
}

@book{Golub,
author = {Golub, Gene H. and Van Loan, Charles F.},
title = {Matrix Computations},
publisher = {Johns Hopkins University Press},
year = {2013},
doi = {10.1137/1.9781421407944},
address = {Philadelphia, PA},
edition   = {4}
}

@article{Ozaki,
  title={{Error-free transformations of matrix multiplication by using fast routines of matrix multiplication and its applications}},
  author={Ozaki, Katsuhisa and Ogita, Takeshi and Oishi, Shin’ichi and Rump, Siegfried M.},
  journal={Numerical Algorithms},
  volume={59},
  number={1},
  pages={95--118},
  year={2012},
  doi={10.1007/s11075-011-9478-1}
}

@Book{Helgaker,
  author =       {T. Helgaker and P. J{\o}rgensen and J. Olsen},
  title =        {{Molecular Electronic Structure Theory}},
  publisher =    {John Wiley \& Sons, Ltd},
  year =         {2000},
  address =      {Chichester},
  DOI={10.1002/9781119019572}
}

@article{Lowdin_Ortho,
author={Per-Olof L{\"o}wdin},
year=1970,
title={{On the nonorthogonality problem}},
journal={Advances in Quantum Chemistry},
volume=5,
pages={185--199},
doi={10.1016/S0065-3276(08)60339-1}}

@article{Carlson,
  title = {{Orthogonalization Procedures and the Localization of Wannier Functions}},
  author = {Carlson, B. C. and Keller, Joseph M.},
  journal = {Phys. Rev.},
  volume = {105},
  issue = {1},
  pages = {102--103},
  numpages = {0},
  year = {1957},
  month = {Jan},
  publisher = {American Physical Society},
  doi = {10.1103/PhysRev.105.102},
}

@article{AAA,
author = {Nakatsukasa, Yuji and S\`{e}te, Olivier and Trefethen, Lloyd N.},
title = {{The AAA Algorithm for Rational Approximation}},
journal = {SIAM J. Sci. Comput.},
volume = {40},
number = {3},
pages = {A1494-A1522},
year = {2018},
doi = {10.1137/16M1106122}
}

@article{mpfr,
author = {Fousse, Laurent and Hanrot, Guillaume and Lef\`{e}vre, Vincent and P\'{e}lissier, Patrick and Zimmermann, Paul},
title = {{MPFR: A multiple-precision binary floating-point library with correct rounding}},
year = {2007},
issue_date = {June 2007},
publisher = {Association for Computing Machinery},
address = {New York, NY, USA},
volume = {33},
number = {2},
issn = {0098-3500},
doi = {10.1145/1236463.1236468},
month = jun,
pages = {13–es},
numpages = {15},
}

@book{gsl,
  title={{GNU scientific library}},
  author={Galassi, Mark and Davies, Jim and Theiler, James and Gough, Brian and Jungman, Gerard and Alken, Patrick and Booth, Michael and Rossi, Fabrice and Ulerich, Rhys},
  year={2002},
  publisher={Network Theory Limited Godalming}
}

@article{Heisenberg_ZfP1934,
author={Heisenberg, W.},
title={{Bemerkungen zur Diracschen Theorie des Positrons}},
journal={Z. Phys.},
volume=90,
pages={209--231},
year=1934,
doi={10.1007/BF01333516},
note={An English translation is found in Arthur I. Miller, \emph{Early Quantum Electrodynamics: A Source Book} (Cambridge University Press, 1995), pp. 169--187},
related={Heisenberg_ZfP1934err},
relatedstring={Corrected in}}

@article{Heisenberg_ZfP1934err,
author={Heisenberg, W.},
title={{Berichtigung zur der Arbeit: Bemerkungen zur Diracschen Theorie des Positrons}},
journal={	Z. Phys.},
year={1934},
month={Sep},
day={01},
volume={92},
number={9},
pages={692-692},
issn={0044-3328},
doi={10.1007/BF01340782},
}

@Article{Dirac:1934:DID,
  author =       "P. A. M. Dirac",
  title =        {{Discussion of the infinite distribution of electrons
                 in the theory of the positron}},
  journal =      {Math. Proc. Camb. Philos. Soc. },
  volume =       "30",
  number =       "2",
  pages =        "150--163",
  year =         "1934",
  DOI =          "10.1017/S030500410001656X",
  fjournal =     "Mathematical proceedings of the Cambridge
                 Philosophical Society",
  received =     "2 February 1934",
}

@InProceedings{Dirac:1934:TDP,
  author =       "P. A. M. Dirac",
  title =        {{Th{\'e}orie du positron. ({French}) [{Theory} of the
                 positron]}},
  booktitle =    {{Structure et propri{\'e}t{\'e}s des noyaux atomiques. Rapports et discussions du septieme conseil de physique tenu {\`a} Bruxelles du 22 au 29 octobre 1933 sous les auspices de l'institut international de physique Solvay. Publiés par la commission administrative de l'institut.}},	 
  editor =       "Institut International de Physique Solvay",
  pages =        "203--230",
  year =         "1934",
  note =         {{Available from \url{https://gallica.bnf.fr/ark:/12148/bpt6k5696894m}}},
  bibsource =    "http://www.math.utah.edu/pub/bibnet/authors/d/dirac-p-a-m.bib",
  language =     "French",
}

@manual{mpmath,
  key     = {mpmath},
  author  = {{The mpmath development team}},
  title   = {mpmath: a {P}ython library for arbitrary-precision floating-point arithmetic (version 1.3.0)},
  url    = { http://mpmath.org/},
  year    = {2023},
}

@inproceedings{RuedenbergRaffenetti,
  note={{Proceedings of the 1972 Boulder Seminar Research Conference on Theoretical Chemistry}},
  author={Ruedenberg, K and Raffenetti, RC and Bardo, RD},
  booktitle={{Energy, structure, and reactivity}},
  year={1972},
  publisher={Wiley},
  address={New York},
  editor={Darwin W. Smith and Walter B. McRae},
  pages={164--169}
}

@article{Visscher1997,
title = {{Dirac–Fock Atomic Electronic Structure Calculations using Different Nuclear Charge Distributions}},
journal = {Atomic Data and Nuclear Data Tables},
volume = {67},
number = {2},
pages = {207-224},
year = {1997},
issn = {0092-640X},
doi = {10.1006/adnd.1997.0751},
author = {L. Visscher and K.G. Dyall}
}

@article{Hamm_JPhysA1990,
	doi = {10.1088/0305-4470/23/17/026},
	year = 1990,
	month = {sep},
	publisher = {{IOP} Publishing},
	volume = {23},
	number = {17},
	pages = {3969--3982},
	author = {A Hamm and D Schutte},
	title = {{How to remove diverges from the {QED}-Hartree approximation}},
	journal = {Journal of Physics A: Mathematical and General},
}

@mastersthesis{Hamm:ddiplomarbeit,
author = {Andreas Hamm},
title= {{Selbstkonsistente Hartree-Korrekturen zur Vakuumpolarisation}},
school = {Universit{\"a}t Bonn},
year = {1988},
type={Diplomarbeit},
note={{An English translation is available from the present authors.}}
}

@article{saue_2006talk,
  author       = {Saue, Trond},
  title        = {{On the variational inclusion of vacuum polarization in 4-component relativistic molecular calculations}},
  month        = nov,
  year         = 2006,
  journal      = {Zenodo},
  doi          = {10.5281/zenodo.4700659},
  note={{Lecture given at the International Conference of Computational Methods in Sciences and Engineering (ICCMSE 2006), symposium \textit{Relativistic quantum theory: Computational perspectives and applications}, held at Chania, Crete, Oct 31 2006}}
}

@article{Kramers_1937,
author={H. A. Kramers},
title={{The use of charge-conjugated wave-functions in the hole-theory of the electron}},
journal={Proc. Roy. Acad. Amsterdam},
volume={40},
year=1937,
pages=814,
note={available from \url{https://www.dwc.knaw.nl/DL/publications/PU00017118.pdf}}}

@Article{Salman_sym2020,
author = {Salman, Maen and Saue, Trond},
title = {{Charge Conjugation Symmetry in the Finite Basis Approximation of the Dirac Equation}},
journal = {Symmetry},
volume = {12},
year = {2020},
number = {7},
article-number = {1121},
ISSN = {2073-8994},
doi = {10.3390/sym12071121}
}

@article{Furry_PhysRev.51.125,
  title = {{A Symmetry Theorem in the Positron Theory}},
  author = {Furry, W. H.},
  journal = {Phys. Rev.},
  volume = {51},
  issue = {2},
  pages = {125--129},
  year = {1937},
  month = {Jan},
  doi = {10.1103/PhysRev.51.125},
  publisher = {American Physical Society}
}

@Article{GrantQuiney2022atoms,
author = {Grant, Ian and Quiney, Harry},
title = {{GRASP: The Future?}},
journal = {Atoms},
volume = {10},
year = {2022},
number = {4},
article-number = {108},
ISSN = {2218-2004},
doi = {10.3390/atoms10040108}
}

@article{Safronova_RevModPhys.90.025008,
  title = {{Search for new physics with atoms and molecules}},
  author = {Safronova, M. S. and Budker, D. and DeMille, D. and Kimball, Derek F. Jackson and Derevianko, A. and Clark, Charles W.},
  journal = {Rev. Mod. Phys.},
  volume = {90},
  issue = {2},
  pages = {025008},
  numpages = {106},
  year = {2018},
  month = {Jun},
  publisher = {American Physical Society},
  doi = {10.1103/RevModPhys.90.025008},
}

@article{sun_TCA2011,
author={Qiming Sun and Wenjian Liu and Werner Kutzelnigg},
title={{Comparison of restricted, unrestricted, inverse, and dual kinetic
balances for four-component relativistic calculations}},
journal=TCA2,
year=2011,
volume=129,
doi={10.1007/s00214-010-0876-6},
pages={423--436}}

@article{Boys_RSPA1950,
author = {Boys, S. F.},
title = {{Electronic wave functions - I. A general method of calculation for the stationary states of any molecular system}},
journal = {Proc. R. Soc. A},
volume = {200},
number = {1063},
pages = {542-554},
year = {1950},
doi = {10.1098/rspa.1950.0036},
}

@book{LAPACK99,
  author    = {Anderson, E. and Bai, Z. and Bischof, C. and Blackford, S. and
               Demmel, J. and Dongarra, J. and Du Croz, J. and Greenbaum, A. and
               Hammarling, S. and McKenney, A. and Sorensen, D.},
  title     = {{LAPACK Users' Guide}},
  edition   = {Third},
  year      = {1999},
  publisher = {Society for Industrial and Applied Mathematics},
  address   = {Philadelphia, PA},
  isbn      = {0-89871-447-8 (paperback)}
}

@article{Reeves_JCP1963a,
    author = {Reeves, C. M.},
    title = {{Use of Gaussian Functions in the Calculation of Wavefunctions for Small Molecules. I. Preliminary Investigations}},
    journal = {The Journal of Chemical Physics},
    volume = {39},
    number = {1},
    pages = {1-10},
    year = {1963},
    month = {07},
    issn = {0021-9606},
    doi = {10.1063/1.1733982},
}

@article{Reeves_JCP1963b,
    author = {Reeves, C. M. and Harrison, M. C.},
    title = {{Use of Gaussian Functions in the Calculation of Wavefunctions for Small Molecules. II. The Ammonia Molecule}},
    journal = {The Journal of Chemical Physics},
    volume = {39},
    number = {1},
    pages = {11-17},
    year = {1963},
    month = {07},
    issn = {0021-9606},
    doi = {10.1063/1.1733984},
}

@article{Klahn_JCP1985,
    author = {Klahn, Bruno},
    title = {{A generalization of the M{\"u}ntz--Sz{\'a}sz theorem to floating exponents with applications to Gauss- and Slater-type functions}},
    journal = {The Journal of Chemical Physics},
    volume = {83},
    number = {11},
    pages = {5749-5753},
    year = {1985},
    month = {12},
    issn = {0021-9606},
    doi = {10.1063/1.449651},
}

@Article{Mohallem_ZfPD1986,
author={Mohallem, Jos{\'e} R.},
title={{A further study on the discretisation of the Griffin--Hill--Wheeler equation}},
journal={Zeitschrift f{\"u}r Physik D Atoms, Molecules and Clusters},
year={1986},
month={Dec},
day={01},
volume={3},
number={4},
pages={339-344},
issn={1431-5866},
doi={10.1007/BF01437189},
}

@article{Hill_PhysRev.89.1102,
  title = {{Nuclear Constitution and the Interpretation of Fission Phenomena}},
  author = {Hill, David Lawrence and Wheeler, John Archibald},
  journal = {Phys. Rev.},
  volume = {89},
  issue = {5},
  pages = {1102--1145},
  numpages = {0},
  year = {1953},
  month = {Mar},
  publisher = {American Physical Society},
  doi = {10.1103/PhysRev.89.1102},
}

@article{Griffin_PhysRev.108.311,
  title = {{Collective Motions in Nuclei by the Method of Generator Coordinates}},
  author = {Griffin, James J. and Wheeler, John A.},
  journal = {Phys. Rev.},
  volume = {108},
  issue = {2},
  pages = {311--327},
  numpages = {0},
  year = {1957},
  month = {Oct},
  publisher = {American Physical Society},
  doi = {10.1103/PhysRev.108.311},
}

@article{Mohallem_IJQC1986,
author = {Mohallem, J. R. and Dreizler, R. M. and Trsic, M.},
title = {{A Griffin--Hill--Wheeler version of the Hartree–Fock equations}},
journal = {International Journal of Quantum Chemistry},
volume = {30},
number = {S20},
pages = {45-55},
doi = {10.1002/qua.560300707},
year = {1986}
}

@article{Biedenharn_PhysRev.126.845,
  title = {{Remarks on the Relativistic Kepler Problem}},
  author = {Biedenharn, L. C.},
  journal = {Phys. Rev.},
  volume = {126},
  issue = {2},
  pages = {845--851},
  numpages = {0},
  year = {1962},
  month = {Apr},
  publisher = {American Physical Society},
  doi = {10.1103/PhysRev.126.845},
}

@article{Borie_RevModPhys.54.67,
  title = {{The energy levels of muonic atoms}},
  author = {Borie, E. and Rinker, G. A.},
  journal = {Rev. Mod. Phys.},
  volume = {54},
  issue = {1},
  pages = {67--118},
  numpages = {0},
  year = {1982},
  month = {Jan},
  publisher = {American Physical Society},
  doi = {10.1103/RevModPhys.54.67},
}

@Inbook{Eides2007,
author="Eides, Michael I. and Grotch, Howard and Shelyuto, Valery A.",
title={{Lamb Shift in Light Muonic Atoms}},
bookTitle={{Theory of Light Hydrogenic Bound States}},
year="2007",
publisher="Springer Berlin Heidelberg",
address="Berlin, Heidelberg",
pages="131--159",
isbn="978-3-540-45270-6",
doi="10.1007/3-540-45270-2_7",
}

@article{CavusogluSikora2023,
      title={{Impact of the nuclear charge distribution on the g-factors and ground state energies of bound muons}}, 
      author = {\c{C}avu\c{s}o\u{g}lu, Atakan and Sikora, Bastian},
      journal={arXiv preprint arXiv:2311.16855},
      year={2023},
      eprint={2311.16855},
      archivePrefix={arXiv},
      primaryClass={physics.atom-ph},}

@article{Borie1983PhysRevA.28.555,
  title = {{Vacuum polarization corrections and spin-orbit splitting in antiprotonic atoms}},
  author = {Borie, E.},
  journal = {Phys. Rev. A},
  volume = {28},
  issue = {2},
  pages = {555--558},
  numpages = {0},
  year = {1983},
  month = {Aug},
  publisher = {American Physical Society},
  doi = {10.1103/PhysRevA.28.555}
  }

@article{PatkosPachucki2025,
      title={{Antiprotonic atoms with nonperturbative inclusion of vacuum polarization and finite nuclear mass}}, 
      author={Patkóš, V. and Pachucki, K.},
      year={2025},
      journal={arXiv preprint arXiv:2509.07738},
      eprint={2509.07738},
      archivePrefix={arXiv},
      primaryClass={physics.atom-ph}}

@article{Fainshtein_JPhysB1991,
  author={A. G. Fainshtein and N. L. Manakov and A. A. Nekipelov},
  title={{Vacuum polarization by a Coulomb field. Analytical approximation of the polarization potential}},
  journal={Journal of Physics B: Atomic, Molecular and Optical Physics},
  volume={24},
  number={3},
  pages={559},
  year={1991},
  doi={10.1088/0953-4075/24/3/012}
}

@article{Manakov_JETP1989,
author = {N.L. Manakov, A.A. Nekipelov, A.G. Fainshtein},
title={{Vacuum polarization by a strong Coulomb field and its contribution to the spectra of multiply-charged ions}},
journal={JETP},
volume=68,
issue=4,
pages=673,
year=1989,
note={{available from \url{http://jetp.ras.ru/cgi-bin/e/index/r/95/4/p1167}}}}

@article{Manakov_Vestnik2012,
title={{Approximation of the vacuum polarization potential by the Coulomb field}},
author={N.L. Manakov and A.A. Nekipelov},
journal={Bulletin of the Voronezh State University Series: physics, mathematics.},
year={2012},
issue=2,
pages={53-57},
note={{(in Russian; an English translation is available from the present authors)}},
url = {http://www.vestnik.vsu.ru/pdf/physmath/2012/02/2012-02-07.pdf}}

@article{Manakov_Vestnik2013,
title={{Approximation of charge density indiced by a Coulomb field in vacuum}},
author={N.L. Manakov and A.A. Nekipelov},
journal={Bulletin of the Voronezh State University Series: physics, mathematics.},
year={2013},
issue=2,
pages={84-89},
note={{(in Russian; an English translation is available from the present authors)}},
url = {http://www.vestnik.vsu.ru/pdf/physmath/2013/02/2013-02-08.pdf}}

@Inbook{Lewin2014,
author={Lewin, Mathieu
and S{\'e}r{\'e}, {\'E}ric},
editor={Bach, Volker
and Delle Site, Luigi},
title={{Spurious Modes in Dirac Calculations and How to Avoid Them}},
bookTitle={{Many-Electron Approaches in Physics, Chemistry and Mathematics: A Multidisciplinary View}},
year={2014},
publisher={Springer International Publishing},
address={Cham},
pages={31--52},
doi={10.1007/978-3-319-06379-9_2}
}

@article{Gomes2010,
  author  = {Gomes, André S. P. and Dyall, Kenneth G. and Visscher, Lucas},
  title   = {{Relativistic Double-zeta, Triple-zeta, and Quadruple-zeta basis sets for the lanthanides La--Lu}},
  journal = {Theor. Chem. Acc.},
  volume  = {127},
  number  = {4},
  pages   = {369--381},
  year    = {2010},
  month   = {nov},
  doi     = {10.1007/s00214-009-0725-7},
  note    = {{Available from \url{https://doi.org/10.5281/zenodo.7574629}}}
}

@article{Dyall2006,
  author  = {Dyall, Kenneth G.},
  title   = {{Relativistic Quadruple-Zeta and Revised Triple-Zeta and Double-Zeta Basis Sets for the 4p, 5p, and 6p Elements}},
  journal = {Theor. Chem. Acc.},
  volume  = {115},
  number  = {5},
  pages   = {441--447},
  year    = {2006},
  month   = {may},
  doi     = {10.1007/s00214-006-0126-0},
  note    = {{Available from \url{https://doi.org/10.5281/zenodo.7574629}}}  
}

@article{Dyall2011,
  author  = {Dyall, Kenneth G.},
  title   = {{Relativistic Double-zeta, Triple-zeta, and Quadruple-zeta basis sets for the 6d elements Rf--Cn}},
  journal = {Theor. Chem. Acc.},
  volume  = {129},
  number  = {3},
  pages   = {603--613},
  year    = {2011},
  month   = {jun},
  doi     = {10.1007/s00214-011-0906-z},
  note    = {{Available from \url{https://doi.org/10.5281/zenodo.7574629}}}  
}

@article{Almoukhalalati,
    author = {Almoukhalalati, Adel and Knecht, Stefan and Jensen, Hans Jørgen Aa. and Dyall, Kenneth G. and Saue, Trond},
    title = {{Electron correlation within the relativistic no-pair approximation}},
    journal = {J. Chem. Phys.},
    volume = {145},
    number = {7},
    pages = {074104},
    year = {2016},
    month = {08},
    doi = {10.1063/1.4959452}
}

@article{CODATA2022,
    author = {Mohr, Peter J. and Newell, David B. and Taylor, Barry N. and Tiesinga, Eite},
    title = {{CODATA recommended values of the fundamental physical constants: 2022}},
    journal = {Journal of Physical and Chemical Reference Data},
    volume = {54},
    number = {3},
    pages = {033105},
    year = {2025},
    month = {09},
    issn = {0047-2689},
    doi = {10.1063/5.0279860}
}

@article{Thierfelder_PRA2010,
  title = {{Quantum electrodynamic corrections for the valence shell in heavy many-electron atoms}},
  author = {Thierfelder, C. and Schwerdtfeger, P.},
  journal = {Phys. Rev. A},
  volume = {82},
  issue = {6},
  pages = {062503},
  numpages = {10},
  year = {2010},
  publisher = {American Physical Society},
  doi = {10.1103/PhysRevA.82.062503},
}

@article{Soff_PhysRevLett.48.1465,
  title = {Self-Energy of Electrons in Critical Fields},
  author = {Soff, Gerhard and Schl\"uter, Paul and M\"uller, Berndt and Greiner, Walter},
  journal = {Phys. Rev. Lett.},
  volume = {48},
  issue = {21},
  pages = {1465--1468},
  numpages = {0},
  year = {1982},
  month = {May},
  publisher = {American Physical Society},
  doi = {10.1103/PhysRevLett.48.1465},
}

@Article{Dyall_TCA1996,
author={Dyall, Kenneth G. and F{\ae}gri, Knut},
title={{Optimization of Gaussian basis sets for Dirac--Hartree--Fock calculations}},
journal={Theoretica chimica acta},
year={1996},
month={Jul},
day={01},
volume={94},
number={1},
pages={39-51},
issn={1432-2234},
doi={10.1007/BF00190154},
}

@book{Jensen_2007ICC,
  title={{Introduction to Computational Chemistry}},
  author={Frank Jensen},
  isbn={9780470058046},
  year={2007},
  publisher={Wiley}
}

@Article{dyall:kinbal,
  author =       {K. G. Dyall and I. P. Grant and S. Wilson},
  title =          {{Matrix representation of operator products}},
  journal =      {J. Phys. B},
  volume =       {17},
  year =         {1984},
  pages =        {493},
  doi={10.1088/0022-3700/17/4/006},
}

@Article{stanton:kinbal,
  author =       {R. E. Stanton and S. Havriliak},
  title =          {{Kinetic balance: A Partial solution to the problem of
                 variational safety in Dirac calculations}},
  journal =      JCP,
  year =         {1984},
  volume =       {81},
  pages =        {1910},
  doi={10.1063/1.447865}
}

@Article{visscher:kinbal,
  author   =      {L. Visscher and P. J. C. Aerts and O. Visser and W. C.
                   Nieuwpoort},
  title    =      {{Kinetic balance in contracted basis sets for
                  relativistic calculations}},
  doi      =      {10.1002/qua.560400816},
  journal  =      {Int.~J.~Quant.~Chem.: Quant.~Chem.~Symp.},
  volume   =      {25},
  year     =      {1991},
  pages    =      {131},
}

@Article{Dyall_CPL1990,
  author =       {K. G. Dyall and K. F{\ae}gri},
  title =          {{Kinetic balance and variational bounds failure in the
                 solution of the Dirac equation in a finite Gaussian
                 basis set}},
  journal =      {Chem.~Phys.~Lett.},
  volume =       {174},
  year =         {1990},
  pages =        {25},
  doi          = {10.1016/0009-2614(90)85321-3}
}

@article{Dyall_CP2012,
title = {{A question of balance: Kinetic balance for electrons and positrons}},
journal = "Chemical Physics",
volume = "395",
pages = "35 - 43",
year = "2012",
note = "Recent Advances and Applications of Relativistic Quantum Chemistry",
issn = "0301-0104",
doi = {10.1016/j.chemphys.2011.07.009},
author = "Kenneth G. Dyall",
}

@Misc{DIRAC25,
      note = "{DIRAC}, a relativistic ab initio electronic structure program,
      Release {DIRAC25} (2025),
      written by T.~Saue, L.~Visscher, H.~J.~{\relax Aa}.~Jensen, R.~Bast and A.~S.~P.~Gomes,
      with contributions from I.~A.~Aucar, V.~Bakken, J.~Brandejs, C.~Chibueze, J.~Creutzberg, 
      K.~G.~Dyall, S.~Dubillard, U.~Ekstr{\"o}m, E.~Eliav, T.~Enevoldsen, E.~Fa{\ss}hauer, T.~Fleig,
      O.~Fossgaard, K.~G.~Gaul, L.~Halbert, E.~D.~Hedeg{\aa}rd, T.~Helgaker, B.~Helmich--Paris,
      J.~Henriksson, M.~van~Horn, M.~Ilia{\v{s}}, Ch.~R.~Jacob, S.~Knecht, S.~Komorovsk{\'y}, O.~Kullie,
      J.~K.~L{\ae}rdahl, C.~V.~Larsen, Y.~S.~Lee, N.~H.~List, H.~S.~Nataraj, M.~K.~Nayak, P.~Norman,
      A.~Nyvang, G.~Olejniczak, J.~Olsen, J.~M.~H.~Olsen, A.~Papadopoulos, Y.~C.~Park, J.~K.~Pedersen,
      M.~Pernpointner, J.~V.~Pototschnig, R.~di~Remigio, M.~Repisky, C. M. R. Rocha, K.~Ruud,
      P.~Sa{\l}ek, B.~Schimmelpfennig, B.~Senjean, A.~Shee, J.~Sikkema, A.~Sunaga, A.~J.~Thorvaldsen,
      J.~Thyssen, J.~van~Stralen, M.~L.~Vidal, S.~Villaume, O.~Visser, T.~Winther, S.~Yamamoto and 
      X.~Yuan
      (available at \url{https://doi.org/10.5281/zenodo.14833106}, 
      see also  \url{https://www.diracprogram.org})"}

@article{DIRACpaper2020,
author = {T.~Saue and R.~Bast and A.~S.~P.~Gomes  and H.~J.~{\relax Aa}.~Jensen and L.~Visscher and I.~A.~Aucar and R.~di~Remigio and K.~G.~Dyall and E.~Eliav and E.~Fasshauer  and T.~Fleig  and L.~Halbert and E.~D.~Hedeg{\aa}rd  and B.~Helmich-Paris  and M.~Ilia\v{s} and C.~R.~Jacob  and S.~Knecht  and J.~K.~Laerdahl  and M.~L.~Vidal and M.~K.~Nayak  and M.~Olejniczak  and J.~M.~H.~Olsen  and M.~Pernpointner  and B.~Senjean and A.~Shee  and A.~Sunaga and J.N.~P.~van~Stralen},
title = {{The DIRAC code for relativistic molecular calculations}},
journal = {The Journal of Chemical Physics},
volume = {152},
number = {20},
pages = {204104},
year = {2020},
doi = {10.1063/5.0004844},
}

@article{Salomonson_PhysRevA.40.5548,
  title = {{Relativistic all-order pair functions from a discretized single-particle Dirac Hamiltonian}},
  author = {Salomonson, Sten and \"Oster, Per},
  journal = {Phys. Rev. A},
  volume = {40},
  issue = {10},
  pages = {5548--5558},
  numpages = {0},
  year = {1989},
  month = {Nov},
  publisher = {American Physical Society},
  doi = {10.1103/PhysRevA.40.5548},
}

\end{document}